\documentclass[fleqn,usenatbib]{mnras}

\usepackage[T1]{fontenc}
\usepackage{ae,aecompl}

\usepackage{graphicx}   
\usepackage{amsmath}    
\usepackage{amssymb}    

\usepackage[english]{babel}

\usepackage[multi-part-units=single]{siunitx}
\usepackage{multirow,tabularx}
\usepackage[colorinlistoftodos]{todonotes}

\usepackage[pdf]{graphviz}
\usepackage{xspace}
\usepackage{pdflscape}
\usepackage{subcaption}
\usepackage{hyperref}
\usepackage{booktabs}

\usepackage{appendix}

\renewcommand{\v}[1]{\boldsymbol #1}

\newcommand{\hii}{\ensuremath{\mathrm{HII}}\xspace}

\newcommand{\heii}{\ensuremath{\mathrm{HeII}}\xspace}
\newcommand{\heiii}{\ensuremath{\mathrm{HeIII}}\xspace}
\newcommand{\hezd}{\ensuremath{\mathrm{He(II+III)}}\xspace}

\newcommand{\xhii}{\ensuremath{x_\mathrm{HII}}\xspace}
\newcommand{\vmxhii}{\ensuremath{\left<x_\mathrm{HII}\right>_V}\xspace}

\newcommand{\xheii}{\ensuremath{x_\mathrm{HeII}}\xspace}
\newcommand{\xheiii}{\ensuremath{x_\mathrm{HeIII}}\xspace}
\newcommand{\xhezd}{\ensuremath{x_\mathrm{He(II+III)}}\xspace}

\newcommand{\nttt}[2]{\ensuremath{#1 \cdot 10^{#2}}\xspace}

\makeatletter
\if@todonotes@disabled

\else

\fi
\makeatother

\DeclareSIUnit[number-unit-product = {}]\yr{yr}
\DeclareSIUnit[number-unit-product = {}]\Mpc{Mpc}
\DeclareSIUnit[number-unit-product = {}]\cMpc{cMpc}
\DeclareSIUnit[number-unit-product = {}]\ckpc{ckpc}
\DeclareSIUnit[number-unit-product = {}]\Mpch{\si{\per\h\mega\pc}}
\DeclareSIUnit[number-unit-product = {}]\kpch{\si{\per\h\kilo\pc}}
\DeclareSIUnit[number-unit-product = {}]\cMpch{\si{\per\h\cMpc}}
\DeclareSIUnit[number-unit-product = {}]\ckpch{\si{\per\h\ckpc}}
\DeclareSIUnit[number-unit-product = {}]\pc{pc}
\DeclareSIUnit[number-unit-product = {}]\h{\textit{h}}
\DeclareSIUnit[number-unit-product = {}]\c{c}
\DeclareSIUnit[number-unit-product = {}]\dex{dex}
\DeclareSIUnit[number-unit-product = {}]\Msol{{M_\odot}}
\DeclareSIUnit[number-unit-product = {}]\Msolh{\si{\per\h\Msol}}

\defcitealias{eide_epoch_2018}{E18}
\defcitealias{eide_next_2020}{E20}

\title[Opening Reionization]{Opening Reionization: Quantitative Morphology of the Epoch of Reionization and Its Connection to the Cosmic Density Field}

\author[P. Busch, M. B. Eide, B. Ciardi, K. Kakiichi]{
Philipp Busch$^{1,2,3}$\thanks{E-mail: pbusch@mpa-garching.mpg.de}, Marius B. Eide$^{1}$, B. Ciardi$^{1}$ and Koki Kakiichi$^{4}$
\\
$^{1}$Max-Planck-Institut f\"ur Astrophysik, Postfach 1317, D-85741 Garching, Germany\\
$^{2}$ Department of Natural Science, The Open University of Israel, 1 University Road, P. O. Box 808, Raanana 43107, Israel\\
$^{3}$ Department of Physics, The Technion, Haifa 3200003, Israel\\
$^{4}$ Department of Physics, University of California, Santa Barbara, CA 93106, USA
}

\date{Accepted XXX. Received YYY; in original form ZZZ}

\pubyear{2020}

\begin{document}
\label{firstpage}
\pagerange{\pageref{firstpage}--\pageref{lastpage}}
\maketitle

\begin{abstract}
We introduce a versatile and spatially resolved morphological characterisation of binary fields, rooted in the opening transform of mathematical morphology. We subsequently apply it to the thresholded ionization field in simulations of cosmic reionization and study the morphology of ionized regions. We find that an ionized volume element typically resides in an ionized region with radius \SI{\sim8}{\cMpch} at the midpoint of reionization ($z\approx7.5$) and follow the bubble size distribution even beyond the overlap phase. We find that percolation of the fully ionized component sets in when 25\% of the universe is ionized and that the resulting infinite cluster incorporates all ionized regions above \SI{\sim8}{\cMpch}. We also quantify the clustering of ionized regions of varying radius with respect to matter and on small scales detect the formation of superbubbles in the overlap phase. On large scales we quantify the bias values of the centres of ionized and neutral regions of different sizes and not only show that the largest ones at the high-point of reionization  can reach $b\approx 30$, but also that early small ionized regions are positively correlated with matter and large neutral regions and late small ionized regions are heavily anti-biased with respect to matter, down to $b\lesssim-20$.

\end{abstract}




\section{Introduction}\label{sec:intro}

During the epoch of reionization (EOR) the universe undergoes its last phase transition from a predominantly neutral state to almost full ionization. This ionization is driven and maintained by the ionizing photons released from dense structures that are able to reach high enough temperatures for their production. It is as of yet unclear to what degree different types of sources, early stars, stellar remnants, or even quasars \citep{madau_cosmic_2015,kulkarni_evolution_2019,shen_bolometric_2020}, are responsible for the emission of ionizing radiation into the intergalactic medium (IGM).


The primary observational evidence for the occurrence of reionization can be found in the spectra of high-redshift quasars, which clearly show the transition to a fully ionized universe around $z\approx6$ (see \citealt{becker_reionisation_2015} for a review). The same spectra also show a certain variability in the exact time of this transition, indicating an inhomogeneous process, which is characterised by ionized regions located around the sources, often called bubbles, that grow until they overlap and subsequently fill the whole universe (e.g. \citealt{gnedin_reionization_2004}).

The high abundance of neutral hydrogen at higher redshifts prevents optical investigations beyond the end of the EoR. Nevertheless, it is precisely that neutral hydrogen that can be detected by its hyperfine transition at a restframe wavelength of 21cm to characterise its distribution at higher redshifts \citep[see][for a review]{pritchard_21_2012}. To this end, many observational efforts have been undertaken or are underway, ranging from measurements of the global signal \citep[e.g EDGES,][]{bowman_toward_2008}, the statistics of the distribution via the power spectrum (such as LOFAR, \citealt{van_haarlem_lofar_2013}, MWA, \citealt{lonsdale_murchison_2009}, PAPER, \citealt{parsons_new_2014}, and HERA, \citealt{deboer_hydrogen_2017}), to prospective actual mappings of the signal \citep[e.g SKA,][]{mellema_hi_2015}.


As the process of reionization is highly non-linear and inhomogeneous, the model predictions can be tested comparing different quantities from mock and real observations as they become available. These models are usually evaluated with semi-analytic \citep{choudhury_semi-analytic_2001,choudhury_updating_2006}, semi-numeric \citep{mesinger_21cmfast_2011,xu_islandfast_2017,greig_21cmmc_2018} or fully numerical simulations \citep{ciardi_simulating_2003,mellema_c2-ray_2006,ciardi_effect_2012,ross_simulating_2017,eide_epoch_2018,ghara_prediction_2018}, sometimes even directly coupled to hydrodynamics and gravitational structure formation \citep{gnedin_cosmic_2014,ocvirk_cosmic_2016,pawlik_aurora_2017,rosdahl_sphinx_2018}. While semi-analytic methods have to compromise on accuracy, they are fast enough for parameter space explorations that are prohibitively expensive using full radiative transfer simulations and prohibitive in the case of coupled simulations.


Even on their own the global 21cm signal and its power spectrum are powerful probes that provide abundant information about the progress of reionization \citep{furlanetto_characteristic_2006,greig_global_2017}. Unfortunately the former does not tell us anything about its inhomogeneity and the latter only provides an exhaustive statistical description if the 21cm signal was a Gaussian random field, which by all expectations is far from the truth \citep{pritchard_21_2012}. Although modern power spectrum based applications yield a large amount of information \citep{shimabukuro_beyond_2020}, it is therefore necessary to find other, more suited statistical descriptions of the morphology of ionization during the EoR.

Unlike the ionization field, the matter overdensity on large scales is very close to a Gaussian random field. As the matter density is notoriously hard to investigate directly, it is interesting to find a statistical connection to the more accessible tracers, usually galaxies, here IGM properties. One such connection is the bias of the tracer density field (in our case neutral patches or ionized bubbles) with respect to the overdensity field \citep{desjacques_large-scale_2018}. While this bias has been investigated on the field level \citep[e.g][]{shin_cosmological_2008,xu_h_2019} by cross-correlating HI with matter, we will extend this to neutral and ionized regions of a given size.

The characterisation of the morphology of reionization has been developed along with the increasing capabilities to simulate the process. Sometimes this morphology traces the 21cm signal \citep{wang_topology_2015} and others ionization fractions directly \citep{friedrich_topology_2011}. An early method that still sees considerable use today is based upon Minkowski functionals \citep{gleser_morphology_2006,kapahtia_novel_2018,chen_stages_2018,bag_shape_2018}. While these are mathematically well developed, they have the disadvantage of being integral quantities of a given surface and are therefore inherently non-local. While this does not pose a problem in the case of isolated, monolithic bubbles, it does so once overlap, or even percolation, sets in. We would therefore like to extend previous approaches to separate connected ionized regions into separate bubbles to carry the notion of bubbles over to the overlap phase, such as \cite{lin_distribution_2016}, by joining them with the concept of granulometry \citep{kakiichi_recovering_2017}.


A problem from an observational point of view is the presence of noise that can hamper morphology detection \citep{kakiichi_recovering_2017}. Possible ways to mitigate it are smoothing the image or introducing super-pixels \citep{giri_optimal_2018}. We neglect a special treatment of this problem for now, deferring it to a future publication, but prefer super-pixels for a real mock observation application. 


In \autoref{sec:sims} we give a short overview of the simulations used in this paper and first presented in \cite{eide_epoch_2018}. In the following we will introduce our new methodology for describing the morphology of ionized regions, both locally and globally, which in \autoref{sec:methods} extends granulometry as introduced by \cite{kakiichi_recovering_2017}. How this method can be used to define a discrete representation of the bubble structure is shown in \autoref{sec:disc_rep}. The results of our analysis are split in three sections: the global properties of the bubble population (\autoref{sec:opening}), the percolation transition (\autoref{sec:percolation}) and finally the connection to the density and luminosity fields in \autoref{sec:physfields}.
 
\section{The Simulations}\label{sec:sims}

The simulations of reionization used for this investigation are the product of a multi-frequency radiative transfer post-processing first presented in \citet[][hereafter E18]{eide_epoch_2018} of the MassiveBlack-II cosmological hydrodynamics-simulation \citep[MBII;][]{khandai_massiveblack-ii_2015} using the monte-carlo ray tracing code \verb+CRASH+ \citep{Ciardi2001,maselli_crash_2003,Maselli2009,graziani_crash3_2013,Graziani2018,Glatzle2019}.

The MBII simulates a $(\SI{100}{\cMpch})^3$ periodic volume with a WMAP7 cosmology and samples the matter and gas distribution with $1792^3$ particles each. The gas and dark matter particles have a mass of $m_{\mathrm{gas}}=\nttt{2.2}{6}\si{\Msolh}$ and $m_{\mathrm{DM}}=\nttt{1.1}{7}\si{\Msolh}$, while the Plummer-equivalent softening length is $\epsilon = \SI{1.85}{\ckpch}$. The simulation was performed using the unpublished code \textsc{P-GADGET} ultimately based upon \textsc{GADGET2} \citep{springel_cosmological_2005}. This code uses smoothed particle hydrodynamics to follow the gas dynamics and implements a number of feedback mechanisms.

For post-processing radiative transfer simulations, we mapped the MBII simulation to grids of $256^3$ cells and therefore a comoving sidelength of $l_c = \SI{0.391}{\cMpch}$. For this, the gas and star particles of the MBII are directly binned in the grid cells. Cells with at least one stellar particle represent sources from which photon packets are emitted. As the number of relatively faint sources increases rapidly below $z=10$ we use a spatial clustering approach to group sources in a flux conserving manner (see Eide et al. in prep., hereafter E20). 

The five different source scenarios as listed in \autoref{tab:scenarios} were first presented in \citetalias{eide_epoch_2018}. They are based on stellar, X-ray binary (XRB), shock heated interstellar medium (ISM) and nuclear accreting black hole (BH) emission, and combinations of these. Here we briefly remind the reader of their key aspects and refer to \citetalias{eide_epoch_2018} and E20 for further details. In the analysis in \autoref{sec:opening} and onward we will denote the scenarios with the labels as given in \autoref{tab:scenarios}. 

\begin{table}
  \caption[Reionization Scenarios]{Source scenarios and associated labels.}
  \centering
  \label{tab:scenarios}
  \begin{tabular}{cl}
    \hline
    Label & Source types \\
    \hline
    Stars & Stars only\\
    SBH & Stars and nuclear accreting black holes\\
    SXRB & Stars and X-ray binaries\\
    SISM & Stars and shock heated interstellar medium\\
    SXBI & All sources\\                                        
    \hline
 \end{tabular}
\end{table}

The basic source of ionizing photons in all simulations are stars. Each stellar particle from the parent simulation is emitting a single stellar population spectral energy distribution (SED) without binary stars according to its age and metallicity. The SEDs are then averaged over the whole simulation at a given snapshot to obtain a mean SED, which is assigned to all source cells and rescaled by the luminosity in a given cell. This procedure is also followed for the other emission of essentially stellar origin (XRBs, ISM). As not all source cells contain active BHs, their emission is treated separately. The stellar emission is the dominating component in the global mean galactic SED at energies up to the ionization energy of \heii (\SI{54.4}{\eV}). We refer the reader to Figure 2 in \citetalias{eide_epoch_2018} for a comparison of the SED contributions of the different source types and to Section 2.3 therein for a detailed description of the SED synthesis.

\section{Methodology}\label{sec:methods}

The aim of this paper is the characterization of the topology of ionized regions in the simulations presented above. For this we first need a definition of such regions and furthermore a framework to quantify their shape. We express the ionized regions as objects in sets of binary fields derived from the ionization fields (\autoref{sec:bf_intro}), and characterize their outer and inner structure with the opening field (\autoref{sec:of_intro}) and the euclidean distance transform (\autoref{sec:edt_intro}). We also introduce a bubble definition based on the hierarchy in the opening field that can be described by a tree structure.

\begin{figure*}
  \includegraphics[width=\linewidth]{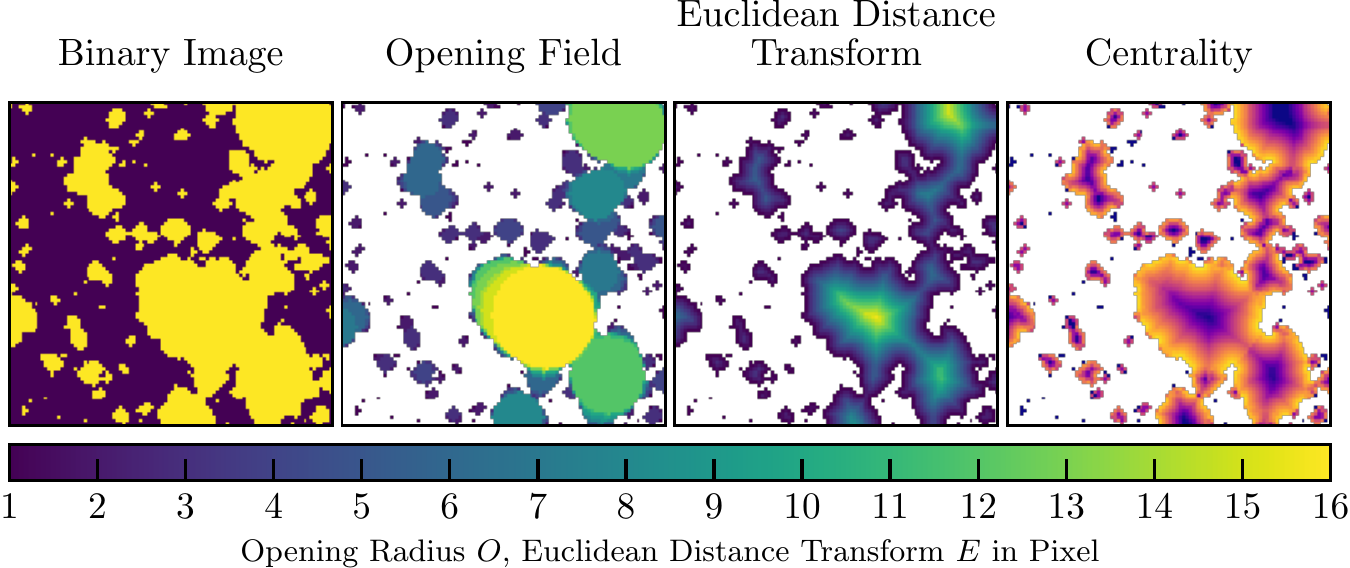}
  \caption{Overview of the concepts introduced in \autoref{sec:methods}: Two-dimensional toy examples of an opened binary image (left panel, see \autoref{sec:bf_intro}) together with the connected opening field (center left panel, see \ref{sec:of_intro}), Euclidean distance transform (center right panel, see \ref{sec:edt_intro}) and centrality (right panel, see \ref{sec:centrality}). The colourbar gives both the value of the opening field and the Euclidean distance. The centrality values vary linearly from 1 (purple) in the most central regions to 0 (yellow) at the very edge.
}\label{fig:field_example}
\end{figure*}

\subsection{Binary Field}\label{sec:bf_intro}

At the core of all tools used in the following to characterize the ionized regions and other fields in the EoR lies the binary field (BF), which is the result of a classification of volume  as either filled (1) or empty (0). 

In the case of ionized regions, we obtain these classifications by comparing the fraction $x_\mathrm{I}(\v r)$ of the ion species $\mathrm{I}$, a field of real values, with a threshold $t_\mathrm{I}$:

\begin{equation}
 X_\mathrm{I}(\v r) = \begin{cases}
            1 & \text{if }x_\mathrm{I}(\v r) \geq t_\mathrm{I},\\
            0 & \text{if }x_\mathrm{I}(\v r) < t_\mathrm{I}.
           \end{cases}
\label{eq:binary}
\end{equation}
Cells are thus considered filled/empty if they lie above or below an ionization threshold. 
It is intuitively clear that for a given field different choices of threshold can result in very different binary fields.

The discretisation of a continuous field on only two values comes with a loss of information in most instances. In cases where the field is almost binary to begin with, such as the ionization fraction obtained with stellar type sources, this merely removes ambiguity. For fields with a large range of values, such as when more energetic sources are considered, one should take sufficiently close thresholds sampling the whole range of values to grasp the possibly changing morphology in different regimes. We will address this point in a companion paper.

\subsection{Opening Field}\label{sec:of_intro}

Next we introduce our morphological classification of the BF.
As regions in the binary field are featureless within their boundaries, the only distinguishing feature between regions is their size and the shape of their surfaces. A simple measure for the size of a region is its diameter, but unfortunately its value heavily depends on the direction along which it is measured. To circumvent this problem we instead find the locally minimal diameter via the morphological opening of the binary field as described in \autoref{sec:of_def}. We then connect this approach to the previously employed technique of granulometry in \autoref{sec:gran_intro}.

\begin{figure*}
  \includegraphics[width=\linewidth]{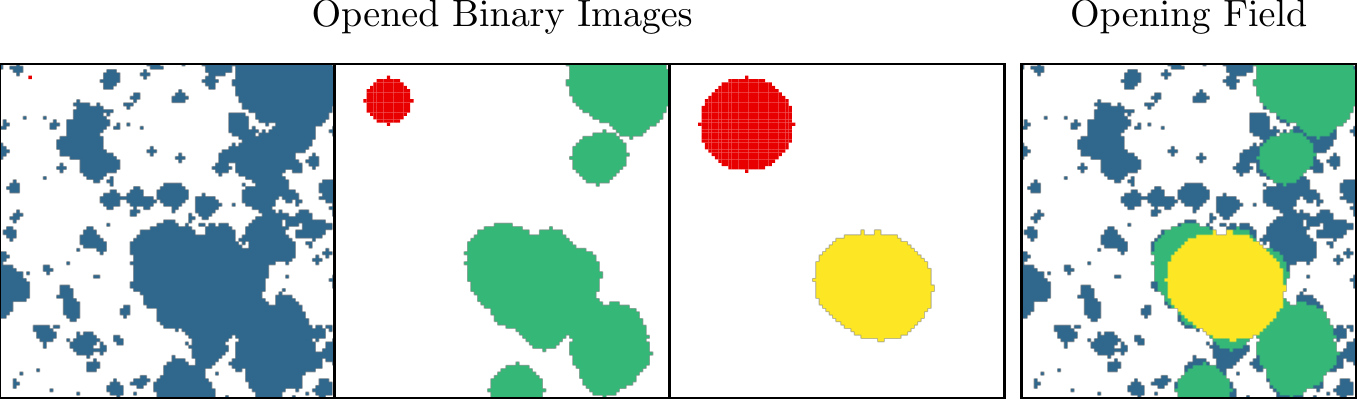}
  \caption{Example of the construction of the opening field in two dimensions. The left three panels show the same binary field morphologically opened with three different structuring elements (SEs) of increasing size from left to right. The corresponding SE is shown in red in the upper left corner of each panel. The rightmost panel shows the resulting opening field constructed from these three opening levels.}\label{fig:opening_const}
\end{figure*}

\subsubsection{Opening Classification}\label{sec:of_def}

The morphological opening can be intuitively understood as the filling of a volume $V \subseteq \mathbb{X}^d$ with overlapping replications of a certain shape, the structuring element (SE) $S \subseteq \mathbb{X}^d$. While the replications can overlap with each other, they must never overlap the boundary. Here $\mathbb{X}$ are usually either the reals $\mathbb{R}$ or, as is the case of our simulation grids, the integers $\mathbb{Z}$.

More formally, the morphological opening is the result of a successive morphological erosion and dilation of $V$ with $S$. The morphological erosion (written as $\ominus$) is the set of all possible centres $\v x \in \mathbb{X}^d$ of translated replications of $S$, here given as $S_{\v x} = \{\v s + \v x| \v s \in S\}$, that are fully contained within $V$:
\begin{equation}
 V\ominus S = \{\v x \in \mathbb{X}^d| S_{\v x} \subseteq V\}.
\end{equation}
For the present application this translates into a removal of a layer of width $r$ around every surface pixel when we erode a given BF with a sphere of this radius.

The morphological dilation (written as $\oplus$) is closely related to the erosion (see also \autoref{sec:four_dil}) and is the union of all $S_{\v x}$ that are translated within $V$:
\begin{equation}
 V\oplus S = \bigcup_{\v x \in V} S_{\v x}.
\end{equation}
The concatenation of the two operations leads to a filtered version $V'$ of $V$ that is the union of all translations of $S$ fully contained within $V$:
\begin{equation}
 V' = V \circ S = (V\ominus S) \oplus S = \bigcup \{S_{\v x}|S_{\v x} \subseteq V\}, \label{eqn:opening_def}
\end{equation}
where $\cdot\circ\cdot$ is the opening operator as explained above. For an alternative description based on Minkowski addition and subtraction see \cite{kakiichi_recovering_2017} and the illustrations therein.

We will exclusively use spheres (or rather their cell-approximations) of increasing radius $R$ as structuring elements to filter the binary field. The spheres are members of a set of binary fields defined by the inequality
\begin{align}
 \|\v x\| \leq R = i\cdot l_c,\text{ for }i\in\mathbb{N}_0,\label{eqn:ball_rad}
\end{align}
where $\v x$ is the position vector to the centre of a cell, $l_c$ is the cell size and $\mathbb{N}_0$ denotes the natural numbers including zero. This implies that the smallest sphere is a single pixel around the origin, the second one is a cross around it and so forth. While the discretisation errors are substantial for small $R$, they decrease quickly with increasing $R$.

While in \cite{kakiichi_recovering_2017} the result of the opening operations was only saved in order to track the evolution of the global volume of the filled regions, here we are also interested in the location of regions of a given local diameter.

An example of the opening of a binary image with disks (the 2D equivalent of spheres) in two dimensions is given in \autoref{fig:opening_const}. The left panel shows the original image (or equivalently opened with a single pixel disk), while the two central panels have been morphologically opened with subsequently larger pixellated disks. 
As the radius of the structuring element used to perform the opening operation on the binary field increases (from left to right), only increasingly larger regions are retained, as those with radii below that of the SE are removed.
From the figure it is clear that the results of an opening with structuring elements of larger $R$ are always contained within those using smaller $R$. 

We exploit this fact by saving only the largest radius at which a given cell is retained after opening a binary image, and name the result of this operation opening field (OF) $O$, i.e. the field of the largest structuring element radius at which a given point in space is still in the opened binary image (see also Table~\ref{tab:radius_defs}). The right-most panel of \autoref{fig:opening_const} shows the three-level opening field constructed from the opened binary images in the other three panels. In the second panel of \autoref{fig:field_example} we show the opening field resulting from opening the binary image of the first panel with structuring elements of radii ranging from 1 to 16 pixels. 
In the simplest case of isolated disks/spheres, the OF just contains these labelled with their radius. In a more realistic scenario, different parts contained within the same ionized region might be removed at different stages of the opening hierarchy described above and thus be assigned a different value of $O$.

While we have so far concentrated our discussion on the ionized component of the gas, it is also of interest to measure the size of its complement. This can be done by simply applying the above methodology to the negation of the initial binary field, i.e. the neutral regions. The result of this operation is saved within the same opening field with a negative sign, but it is not shown in \autoref{fig:field_example}, where for clarity the neutral regions are all colored in white.

\subsubsection{Granulometry}\label{sec:gran_intro}

Granulometry measures the global volume distribution in regions of a binary field with a given diameter. It uses successive openings  with spheres as described in the previous section, and records the volume loss after each step. \cite{kakiichi_recovering_2017} introduced this technique to the field of reionization studies in their investigation of the expected 21cm spot size.

The central quantity in granulometry is the pattern spectrum
\begin{equation}
 F(<R) = 1 - \frac{V(B\circ S_R)}{V(B)},
\end{equation}
where $B$ is the binary image, $V$ is the volume function and $S_R$ the sphere of radius $R$.  $F(<R)$ is thus the cumulative probability distribution of residing in a region of radius smaller than $R$, and the volume weighted probability distribution of being in a region with radius $R$ is 
\begin{equation}
 p_V(R) = \frac{\mathrm{d}F}{\mathrm{d}R}.
\end{equation}

By construction, the opening field from the previous section allows for a quick calculation of the pattern spectrum, as it records the largest radius $R$ of a sphere for which a point is still included in the opening with that sphere. The volume variation between openings with sphere size $R$ and $R+\Delta R$ is therefore just the volume for which $R\leq O \leq R+\Delta R$. We thus just evaluate the volume of all cells contained within a bin of $O$. After normalisation by the total volume contained in the binary image, the bin volume provides the variation in $F$ over that bin, which, divided by the bin width, finally gives a finite difference approximation of the probability density $p_V$ in the bin.

\subsection{Euclidean Distance Transform}\label{sec:edt_intro}

In addition to the opening field from the previous section, to characterize our BF we also use the Euclidean distance transform (EDT), which assigns every filled cell (as defined in eq.~\ref{eq:binary}) the minimal Euclidean distance $d$ to an empty cell. In the context discussed here, the EDT thus measures the distance from the first layer of cells that are outside the ionized regions, hence we will also refer to it as surface distance. The result of this operation for our example images can be seen in the third panel of \autoref{fig:field_example}.

As (filled) spheres in Euclidean space are sets of points with an upper limit on the Euclidean distance from their centre, the EDT can also be seen as the distribution of centres of maximally large spheres contained fully within the filled volume. This highlights the tight connection to the opening field with spherical structuring elements, to which it forms a kind of dual field. The same is true for our pixelated spheres.
In fact, if both fields are needed for a given binary image $B$ it is faster to compute the EDT $E$ and then dilate thresholded versions of it. In the special case of spherical SEs, the opening field $O$ can then be found for every point in space as
\begin{align}
O = \max R_j \left(B \circ S_j\right) = \max R_j |E>R_j| \oplus S_j ,\label{eqn:edt_opening}
\end{align}
where $S_j$ is the structuring element of radius $R_j$ and $|E>R_j|$ is the binary field of regions in $E$ that have a value higher than $R_j$\footnote{For spherical structuring elements the region of the EDT at and above level $R$ is equivalent to the result of an erosion operation with an SE of radius $R$, making Equations \eqref{eqn:edt_opening} and \eqref{eqn:opening_def} equivalent.}.

The EDT field has two major uses: its relation to the OF gives a measure of centrality in highly irregular objects as discussed in \autoref{sec:centrality} and it allows to define distance profiles with respect to the boundary of our objects as described in the following.

As for the opening field before, we also characterize the space between the ionized regions (i.e. the neutral regions) by calculating the EDT of the empty cells (i.e. the negation of the binary image), and save it with a negative sign in the same field as the original positive one. Also in this case, for the sake of clarity, this is not shown in \autoref{fig:field_example}.

\subsection{Centrality}\label{sec:centrality}

A very helpful derived quantity using both the opening field value $O$ and the EDT value $E$ is the centrality $C$, defined as:
\begin{equation}
 C = \frac{E}{O}.
\end{equation}
A two-dimensional example can be found in the rightmost column of \autoref{fig:field_example}. As the OF provides the local radius of the region, this ratio gives a dimensionless measure of how locally central a point is, i.e. how far away from the surface a point is located, given the opening value. This helps to localise features in bubbles in a dimensionless way and to further separate overlapping bubbles. Two important caveats due to discreteness should be mentioned, one associated to the grid representation and one to the steps taken during the opening process. The imperfect discrete representation of spheres on a grid, especially if the radius is only a few cells size, leads to a situation where the central cell has an Euclidean distance to the closest empty cell which is smaller than the opening value. Due to the improving representation with increasing number of cells this error vanishes for well resolved spheres.

Due to these discreteness problems we only approach $C=1$ in the limit of an arbitrarily fine grid and one opening step for every unique value of the EDT.
This optimal spacing of the opening steps is very costly, especially on highly resolved grids, as the number of unique EDT values in 3D grows as $r^{\sim1.9}$ with the radius of the region\footnote{The value of the exponent has been obtained by the authors via direct numerical evaluation for the range of structuring elements used in this work.}. When considering only a coarser set of steps, e.g. only integer multiples of the cell size, the centrality in the centre of regions in many cases will not be exactly unity but slightly smaller. For an integer opening stepping on a grid with spacing $l_c$ every cell with
\begin{equation}
 C>1-\frac{l_c}{O}\label{eqn:centrality}
\end{equation}
will be maximally central as the OF and EDT value can differ by up to, but not including, one cell size before an additional opening step would occur and reduce the difference to zero.

As we adopt integer multiples of the cell size as opening radii (see eq.~\eqref{eqn:ball_rad}), we therefore use this criterion when we identify the centres of bubbles as will be explained in the following section (see Sec.~\ref{sec:bubble_def}).

\section{Discrete Representation of the Bubble Structure}\label{sec:disc_rep}

Our next step is to find an abstraction of the highly structured OF and EDT as introduced in the previous section. More specifically, we want to represent the structure of the binary field by a combination of building blocks, the bubbles, and their connections. We define bubbles as a hierarchy of regions with decreasing opening value, i.e. regions that can be described by nested spheres whose centers lie within a larger sphere (apart from the largest sphere at the root of the hierarchy). The hierarchy is encoded in the ``bubble tree'' representing a given bubble, while the spheres needed for this description are given in the ``minimal centre set''. These are presented in more details in the following.

It should be highlighted that these concepts are fundamentally different from the previously introduced OF and EDT, in that they provide a discretised description of the structure of these fields. Therefore, the morphological properties of the binary field as a whole (e.g. granulometric size distribution and distribution of distances to the ionization front) are best described by the OF and EDT, while its structure (e.g. the structure around a given source and its development into the overlap phase) is more easily accessible by looking at the discrete bubbles.

\subsection{The Minimal Centre Set}\label{sec:min_struc}

\begin{figure}
 \includegraphics[width=\linewidth]{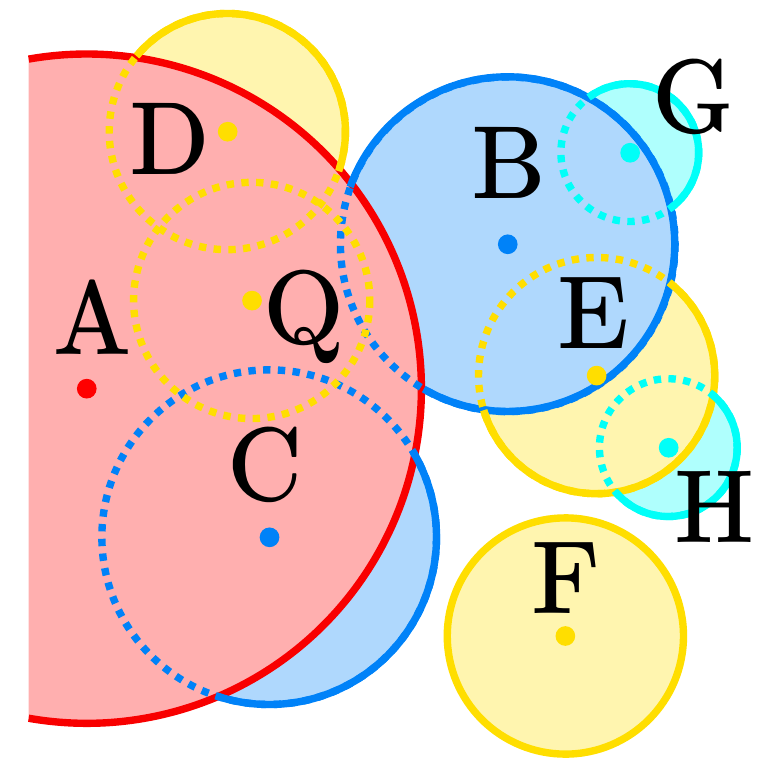}
 \caption{2D schematic of the minimal centre set using disks as structuring elements. The weight of a disk centre (represented by the labelled dots) is calculated as the ratio between the area of the disk which does not belong to a larger disk (shaded regions) and the total surface of that same disk (its boundaries are indicated by solid and dashed circles). Disks with the same opening value are indicated with the same colour.}\label{fig:mcs_weighting}
\end{figure}

\begin{figure}
 \includegraphics[width=\linewidth]{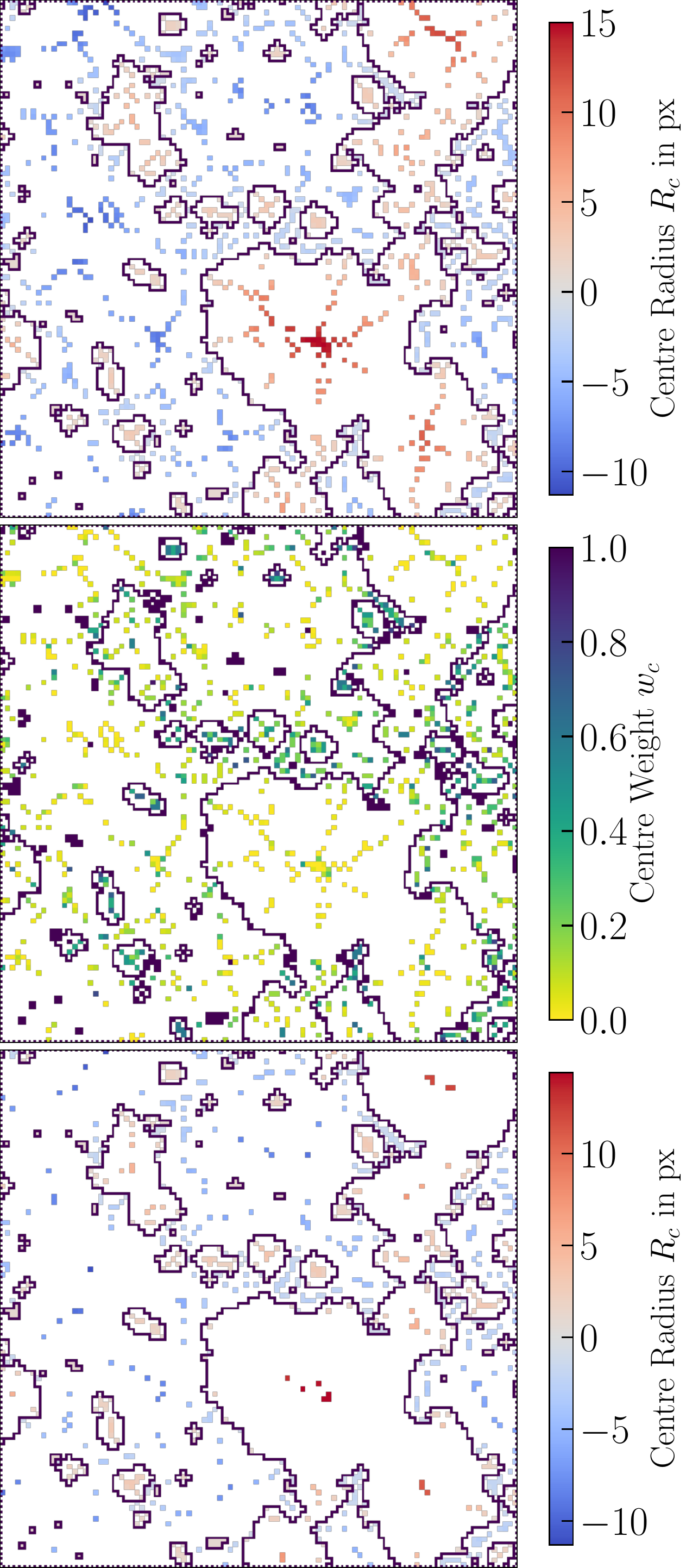}
 
 \caption{Minimal centre set for the 2D example binary field given in \autoref{fig:field_example}. When the BF is an ionization field, the purple contours mark the boundary between neutral and ionized regions. \textbf{Top panel:} radii $R_c$ of the essential structuring elements centred on a given pixel (see text for a definition). Red, positive (blue, negative) values refer to ionized (neutral) disks. \textbf{Middle panel:} weights of the centres of the essential structuring elements. \textbf{Bottom panel:} as the top panel, but only the radii corresponding to a centre with weight $w_c \geq 0.1$ are retained.
 }\label{fig:mcs_example}
\end{figure}

In the representation of the binary field and its accompanying opening field, only a few of the structuring element replications are covering cells that are not contained at a larger opening value. Their centres form the Minimal Centre Set (MCS) and in the remainder we will refer to them as structural centres.

The construction of a MCS in 2D is illustrated in \autoref{fig:mcs_weighting}, where the dots are the pixels identified as centres of ionized disks (the structuring elements) of a given opening value. The MCS will contain only essential structuring elements, i.e. all those disks which can not be omitted without losing information about the shape of the region. Conversely, unessential SEs do not cover a single volume element that is not also covered by a larger SE and thus do not contain useful information. With respect to \autoref{fig:mcs_weighting}, the only depicted unessential SE is disk Q.

The centre of each essential disk is recorded and assigned a weight, $w_c$, proportional to the fraction of the surface which does not belong to a larger disk (indicated by the shaded areas). For example, the largest disk A receives a weight equal to 1. Disks B and C have the same surface, but C receives a smaller weight as it overlaps with A more than B. Similarly, although D and F have the same surface, the latter receives a unit weight as it is fully isolated. 
At the end of this procedure (its 3D equivalent is formally quantified in Appendix~\ref{app:bubble}) we are left with a scalar field of weights that is non-zero only for centres of disks necessary for a reconstruction of the shape of the region they belong to.

The centre radius $R_c$ of a given centre in the MCS is just the value of the EDT at that point (see also Table~\ref{tab:radius_defs}):
\begin{equation}
 R_c(\v x) = E(\v x)\text{, if }w_c > 0.
\end{equation}
This follows from the fact that our spherical SE is confined by the closest point outside of its component in the BF. This minimal distance to the outside is exactly the one measured by the EDT. The OF at the same point will have a larger value if we are not in a local centre (i.e. a point where movement in any direction would bring us closer to the boundary) and we are therefore overlapped by a larger SE. From the convention of treating the OF and EDT values for neutral patches as negative values, follows that we also treat the centre radii of neutral regions as negative (e.g. in \autoref{fig:bias}). Again, this is simply a matter of convenience and does not imply a truly negative radius.

As an example\footnote{To simplify the discussion, the BF will be considered as the product of the thresholding of an ionization field, but the results are of general application.}, we apply the above procedure to the field in \autoref{fig:field_example}, and show the radii of the essential elements and weights of their centres in the top and middle panel of \autoref{fig:mcs_example}, respectively. We find a total of 1343 centres, of which only 814 have  $w_c \geq 0.1$ (see bottom panel of the figure). The vast majority of structuring elements thus removed are tracing shallow irregularities on the surface of large regions, which see a dramatic decrease in the number of structuring element centres within them. This means that the overall structure of the field can be mostly described by the overlap of a relatively small number of disks with large radii, while smaller, more numerous, disks are only needed to reconstruct the relatively small area close to the surface of the ionized region. 

We also notice that the structure of the ionized region (depicted in red) is captured by much fewer elements than the neutral one (depicted in blue; 574 vs 769 in the upper panel, and 357 vs 457 in the lower panel), highlighting their fundamentally different structure: the ionized regions are mostly convex and can therefore be well described by convex shapes like disks, while the neutral region is highly concave and thus necessitates many more essential structuring elements for its characterization. 

By construction, all one-cell structuring elements have a weight of one, but we also find many small disks (a few cells radius) with centres of high weight, indicating the presence of small structures which are not fully embedded within larger disks.
We note that small ionized regions are typically represented by a few small structuring elements of similar size. As such regions grow, surface irregularities on the scale of the grid are much smaller than the region itself and a centre can be more clearly identified. 

\subsection{The Bubble Tree}\label{sec:bubble_def}

\begin{figure}
 \includegraphics[width=\linewidth]{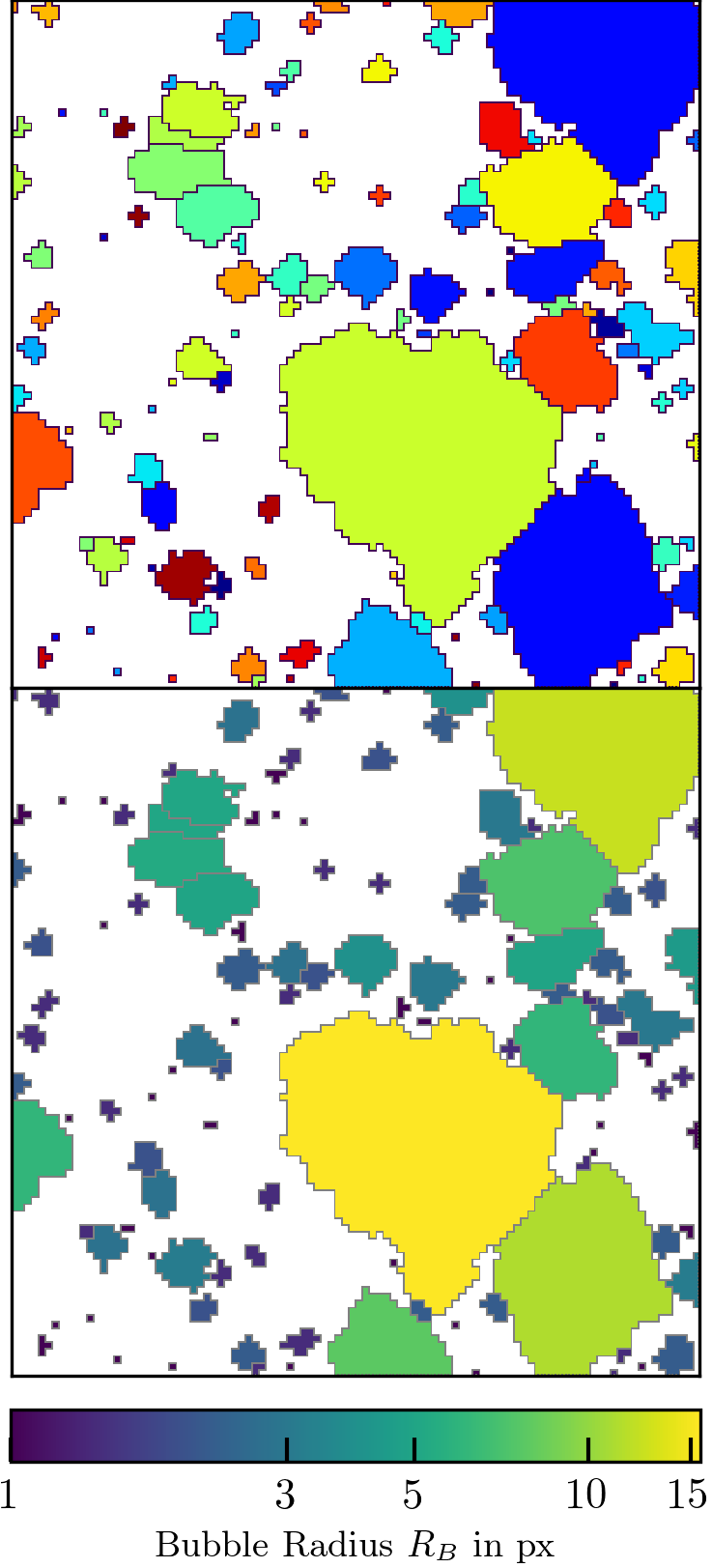}
 
 \caption{Bubble segmentation as described in \autoref{sec:bubble_def} for the example in \autoref{fig:field_example}. \textbf{Upper panel:} Each bubble was given a random colour. \textbf{Lower panel:} The bubbles are coloured by their respective bubble radius $R_B$.}
 \label{fig:bubble_example}
\end{figure}

In the previous section we have been referring to ionized disks and regions, but we still have not provided a definition of an ionized bubble. Here we use the information above to identify an unambiguous, but intuitive definition of a bubble that also captures its inner structure by recording the overlap relationships of the structuring elements contained within it. 
Such a structure constitutes the bubble tree, and it is formed by one root (i.e. a SE whose centre does not belong to a larger one) and its children, i.e. SEs whose centres are covered by larger SEs, the parents. As any given disk/sphere can only have one parent but multiple children (see example below) these relationships constitute a unique tree structure from the largest to the smallest region in any given bubble. As every ionized cell and every centre in the MCS belong exactly to one bubble tree, the construction is unique for a given binary field (the formal procedure is described in \autoref{app:bubble}).

In reference to the 2D example in \autoref{fig:mcs_weighting}, we can identify a bubble formed by disk A with its children C and D, one formed by B, G, E and H (H is a child of E, which in turn together with G is a child of B), and finally the disk with centre F is a bubble in itself and has no children.

We define the centre of the bubbles as the root of the tree, i.e. the point where $C=1-(l_c/O)$ (see \ref{sec:centrality}). Note that in general this point is not necessarily the same as the volume weighted centre of the same bubble. In the simple case that a bubble can be represented by a single structuring element (e.g. object F in \autoref{fig:mcs_weighting}) these centre definitions instead coincide. 

The upper panel of \autoref{fig:bubble_example} is obtained after applying this segmentation procedure to find the bubbles in our 2D example from \autoref{fig:field_example}. Here each bubble is given a random colour to distinguish it from its neighbours. The boundary of each bubble is marked by a black line. Here and in the following applications we impose an additional restriction on the procedure in order to battle effects on the scale of the grid, i.e. we neglect single cell bubbles if they share a face with a bubble of larger radius. Without this restriction we find a very large number of single cell bubbles lining the surface of larger bubbles which most likely are spurious.

We define the radius $R_B$ of a bubble as the maximum value of the EDT field $E$ inside the bubble $B$ (see also Table~\ref{tab:radius_defs}):
\begin{equation}
  R_B = \max_B(E).
\end{equation}
 This simple definition follows from the fact that the largest value of $E$ is found at the root, as otherwise there would be a larger structuring element within the bubble. If this were the case, then by construction of the bubble as a hierarchical set of overlapping SEs, it would have to be the root of the bubble tree. We note that there can be multiple local maxima in the EDT field within a bubble but only the global maximum is not covered by a larger SE and therefore must have the maximum $E$ value. The bubble radii corresponding to the bubbles shown in the upper panel of \autoref{fig:bubble_example} are given in the lower panel of the same figure. 

\begin{table}
  \caption{Radius definitions used throughout this work.}
  \centering
  \label{tab:radius_defs}
  \begin{tabular}{p{21mm}cp{40mm}}
    \toprule
    Name & Symbol & Definition\\
    \midrule
    Opening Radius & $O$ & The radius of the largest structuring element covering a given point in space and being fully contained in one component of the binary field (see \ref{sec:of_def})\\
    \midrule
    Euclidean Distance Transform & $E$ & The distance to the closest point in the complement in the binary field (see \ref{sec:edt_intro})\\
    \midrule
    Center Radius & $R_{c}$ & The radius of the largest structuring element centred on a given point in space and being fully contained in one component of the binary field (see \ref{sec:min_struc})\\
    \midrule
    Bubble Radius & $R_{B}$ & The radius of the largest structuring element at the root of a hierarchy of overlapping structuring elements representing a ``bubble'' (see \ref{sec:bubble_def})\\
    \bottomrule
 \end{tabular}
\end{table}

\section{Opening Analysis of ionization Fields}\label{sec:opening}

After discussing the global trends in the ionization fractions in \autoref{sec:ion_vol} we take a first look at the morphology of the ionization fields of hydrogen and helium. For this we produce binary fields from the 3D ionization fields using a threshold $x_i \geq 0.99$ at redshifts $z\in\{10,9,8,7.5,7,6.5\}$ and apply to them the techniques described in \autoref{sec:methods}. As all aspects of this analysis are connected with the opening field, we will summarize them under the term opening analysis. 

Before discussing our results, it should be noted that the analysis of the helium component of the gas is more complex than that for hydrogen, because of the  appearance of a non-negligible \heiii fraction in the centre of \heii regions. One should therefore use the combined \heii and \heiii fraction $\xhezd=1-x_{\mathrm{HeI}}$. This morphology is virtually the same as the \hii one, for all thresholds, source scenarios and redshifts. In the following we therefore give neither the results for \heii alone nor for \hezd, and defer the discussion for different choices of the threshold to a companion paper.

\begin{figure}
\centering
  \includegraphics[width=\linewidth]{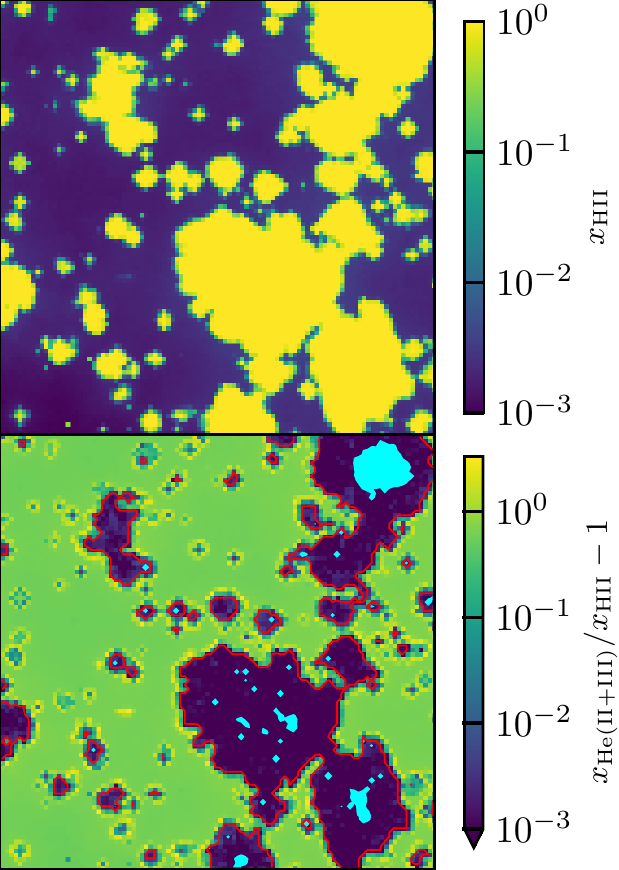}
  \caption{\textbf{Upper panel:} $\left(\SI{39}{\cMpch}\right)^2\times\SI{0.4}{\cMpch}$ example slice of the hydrogen ionization fraction for the complete model (SXBI) at $z=8$. \textbf{Lower panel:} The same slice as above, showing the $\xhii=0.99$ contour in red and the relative difference between \xhii and \xhezd in colour. The aqua coloured patches depict regions in which $\xheiii\geq0.01$ so that $\xheii<0.99$ despite full helium ionization.}\label{fig:ion_frac_example}
\end{figure}

We illustrate the above in \autoref{fig:ion_frac_example}. The upper panel shows the clear dichotomy between fully ionized bubbles, shaped by the relatively soft UV radiation, and the almost neutral regions in between, which are only slightly ionized but heated by more energetic photons. This picture clearly motivates a concentration on the fully ionized regions with $\xhii>0.99$ for a first application of our techniques. In the lower panel we show the absolute relative difference $\xhezd/\xhii-1$ between ionized helium and hydrogen (in colour) in relation to the contour at $\xhii=0.99$ in red. In the regions in which hydrogen is fully ionized it traces ionized helium perfectly. Only in the neutral regions do we see strong deviations between helium and hydrogen ionization. Why it is necessary to compare \xhii to \xhezd instead of \xheii is shown with the aqua-coloured patches, which mark the regions in which $\xheiii>0.01$ and therefore $\xheii<0.99$ despite full helium ionization.

\subsection{Total Ionized Volume}\label{sec:ion_vol}

\begin{figure}
 \includegraphics[width=\linewidth]{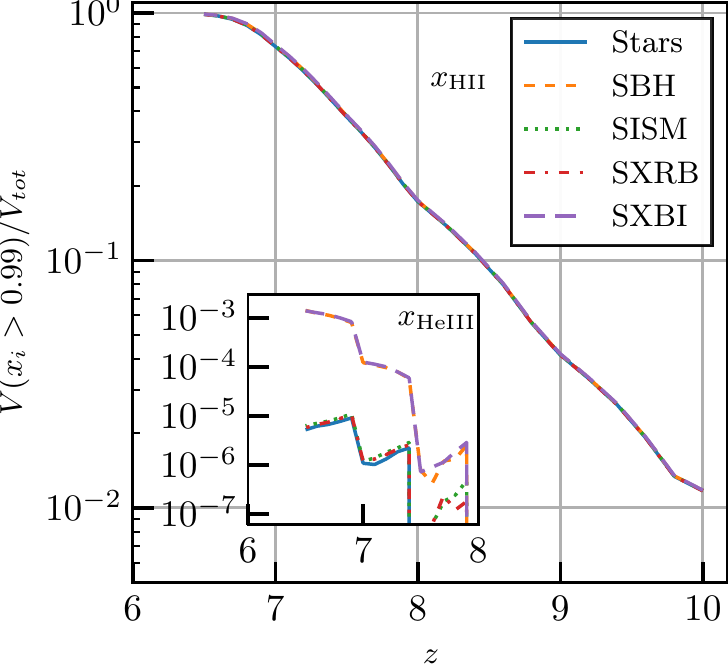}
 \caption{Evolution of the volume fraction of cells with $x_{i}>0.99$ for our simulations sample (indicated by different line styles and colours).}\label{fig:ion_vol}
\end{figure}

We start by showing in \autoref{fig:ion_vol} the fraction of the total volume which is ionized according to our classification, i.e. with $x_i > 0.99$. For simplicity, we will also refer to these cells as fully ionized.

As expected, the fraction of fully ionized cells increases with decreasing redshift for all species, as reionization proceeds. While the evolution of the full ionization of \hii is mainly driven by stellar type sources, and thus is similar for all scenarios, this is not the case for helium.
In fact, from the inset in the figure we see that only the hard spectrum of BHs is efficient at fully ionizing helium and producing an appreciable number of cells with $\xheiii>0.99$, while XRBs and the ISM can only induce a lower level of ionization (see also E20). With BHs, cells start crossing the $\xheiii>0.99$ threshold at $z=7.5$, while without BHs this happens at $z=7$.

\subsection{Typical Dimension of Ionized Regions}\label{sec:gran_res}

To investigate the typical radii of the ionized regions, we apply the granulometric analysis introduced in Section~\ref{sec:gran_intro} to the binary images obtained from our simulations, as described previously. Here we use the full distribution of opening radii, in contrast to the bubble radius, which is the maximum radius in the hierarchy of scales that constitute a bubble (see \ref{sec:bubble_def}), whose distribution is instead investigated in \ref{sec:bnum_res}.

\begin{figure}
 \includegraphics[width=\linewidth]{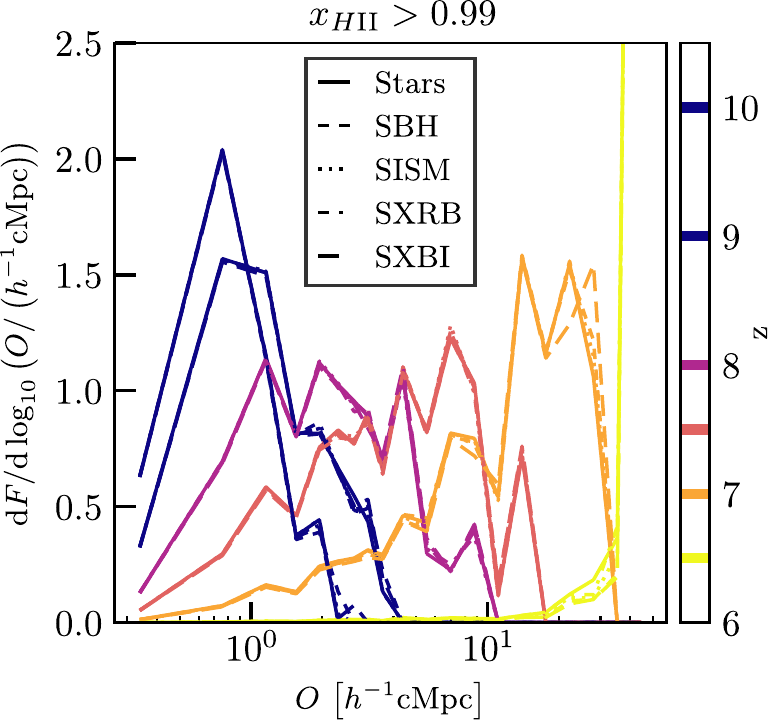}\vspace{0.5cm}
 \includegraphics[width=\linewidth]{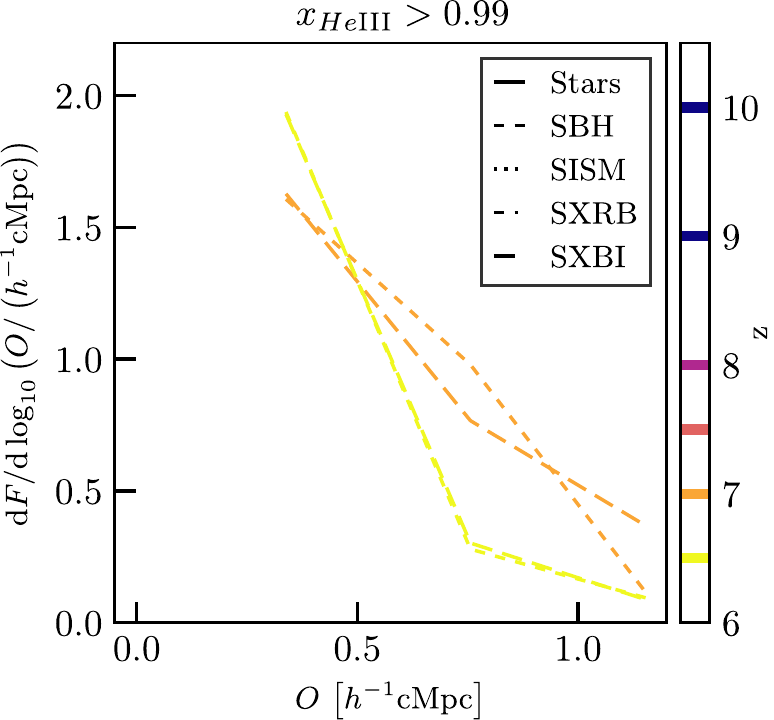}
 \caption[Granulometry Results of \hii \& \heiii Regions]{Volume weighted probability distribution of being in a region of fully ionized (i.e. with ionization fraction $>0.99$) H (top panel) and He (bottom panel) with opening value $O$. Line style refers to different source types, while colour to redshift.}\label{fig:gran_z}
\end{figure}

In \autoref{fig:gran_z} we show the results of the granulometric analysis in terms of the volume weighted probability distribution of being in a fully H ionized region
with opening value $O$. We do not explicitly show the $\xhezd$ component as the same results apply to it. The most striking feature is the almost complete independence from the source scenario. While we see minute differences at the largest radii at any given redshift, it is nonetheless clear that not just the global volume, but also the spatial distribution of cells with high \xhii values is, as expected, completely set by stellar sources. Quantitatively we find a strong increase in the typical opening value in the fully ionized volume from less than \SI{1}{\cMpch} at $z=10$ to about \SI{20}{\cMpch} at $z=7$.  At $z=6.5$ almost all cells are ionized and therefore most of the volume in the simulation box resides at the largest possible opening value of \SI{50}{\cMpch}.
We interpret this as a signature of the increasing overlap of bubbles into a single ionized region. We would like to note that the expected typical size of $\sim\SI{8}{\cMpch}$ at $z=7.5$ is consistent with predictions by \cite{wyithe_characteristic_2004} and the simulation results presented in \cite{iliev_simulating_2006} and \cite{zahn_comparison_2011}. It also is close to the inferred size of a reionization bubble from the recent observations by \cite{tilvi_onset_2020}.

As is to be expected, the \heiii bubble dimensions are much smaller and generally dominated by one-cell regions. There are indeed no cells passing the threshold at $z>7.5$. As discussed in \autoref{sec:ion_vol} we find a crucial dependence on BH sources in the formation of \heiii, as the galactic emission (stars, ISM, XRBs) is not able to carve out bubbles with diameters larger than one cell ($\sim\SI{400}{\ckpch}$).
The rare, hard and powerful BH sources, on the other hand, form bubbles with radii of up to $\approx\SI{1}{\Mpch}$. 
Interestingly, we find a decrease in the typical bubble radius at $z=6.5$ compared to $z=7$. This is the effect of the larger number of BHs that increases the total volume above threshold, but at the same time reduces the average bubble size as typically a newly formed black hole sits in a \heiii region smaller than an old one. 

\subsection{Bubble Numbers}\label{sec:bnum_res}

  \begin{figure}
      \includegraphics[width=\linewidth]{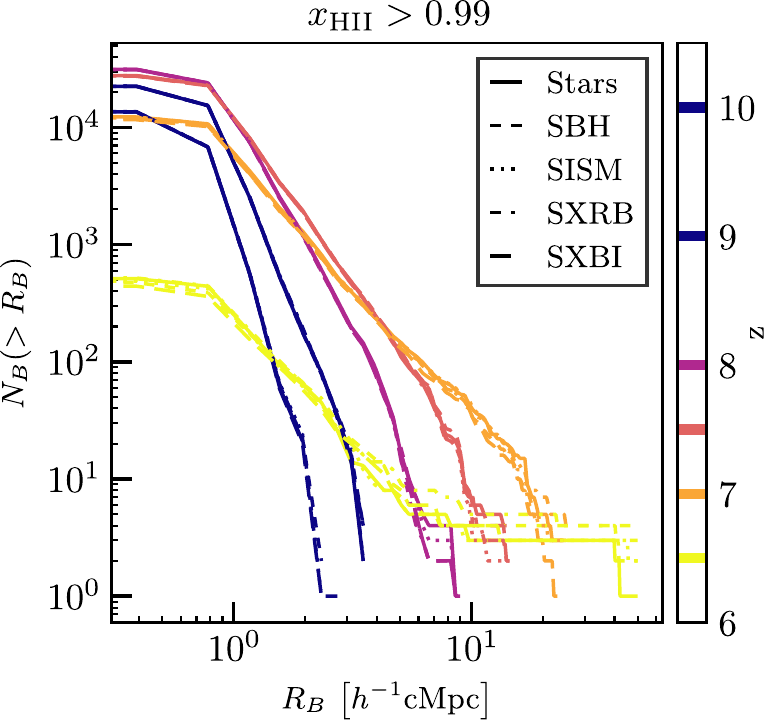}\vspace{0.5cm}
      \includegraphics[width=\linewidth]{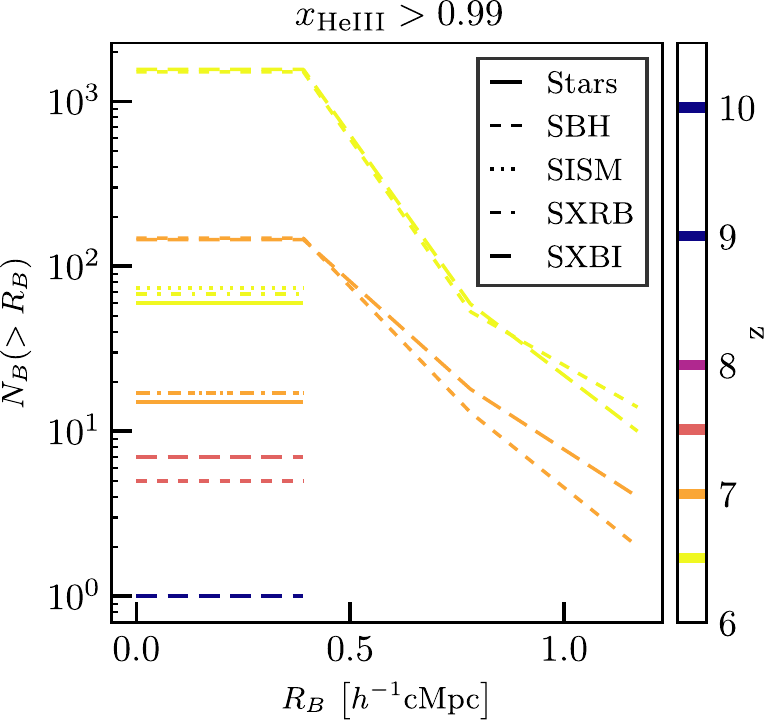}
    \caption[Bubble Number Function]{Number of bubbles with radius largen than $R_B$. Line style refers to different source types and colour to redshift. Top and bottom panel refers to \hii and \heiii regions, respectively.}\label{fig:number}
  \end{figure}

We now investigate the distribution of ionized bubbles by showing in Figure~\ref{fig:number} their number, $N_B$, as a function of the bubble radius, $R_B$, i.e. the maximum radius in a bubble as described in Section~\ref{sec:bubble_def}. $R_B$ is the characteristic scale of a bubble and coincides with the single radius of a perfectly spherical one.

As done in the previous section, we only show the results for \hii and \heiii.
As expected from the luminosity distribution in a cosmological context, the bubble number distribution in \hii at $z>7$ follows a truncated power law, with a slope which becomes shallower for decreasing $z$, especially below $z=8$. This is also the time when the maximum number of independent bubbles is reached ($N_B\gtrsim\nttt{3}{4}$), before further overlap and mergers of bubbles decrease their abundance again, resulting in a similar number of bubbles at $z=10$ and $z=7$. Although there are minute differences at the large end of the distributions, the numbers are virtually identical for all scenarios of reionization.
The distribution at $z=6.5$ follows the same slope as that at $z=7$ for small radii (i.e. it is shaped by the luminosity function), which extends to a long tail at larger radii without exhibiting any truncation. This is associated to the formation of a few extremely large bubbles which cover almost the full simulation volume, suggesting that larger boxes would be needed to properly sample these scales. We would, in fact, expect a truncation similar to the one observed at earlier times. 

For \heiii the situation is again very simple, with a non negligible number of extended \heiii regions appearing only in scenarios involving BHs. We find the first $\SI{1.2}{\cMpch}$ ionized regions at $z=7$, when there are already over 100 bubbles, which become more than 1,000 at $z=6.5$. Their size, though, remains limited to $R_B \sim \SI{1}{\cMpch}$. The differences between the SBH and SXBI scenarios are on the levels of a few bubbles and hence not significant. When BHs are not present, we find less than 20 (100) single cell bubbles at $z=7$ ($6.5$). 

\section{Percolation Analysis}\label{sec:percolation}

As the ionized bubbles grow, they start to overlap into larger regions, eventually leading to the formation, through a percolation process, of a connected structure spanning the whole box (i.e. touching every face of the simulation box), which we refer to as percolating object (PO).
We define the onset of percolation as the time when connections\footnote{Two cells are considered connected if they share a face.} down to single cells are needed to form a cluster spanning the entirety of the simulation box in all three dimensions. Due to the missing periodicity of the box we do not consider the wrapped connection of the regions.
We note that, while the percolation results of \cite{furlanetto_reionization_2016} agree  qualitatively with ours, it is difficult to make a quantitative comparison due to the very different approaches and resolutions used in the underlying simulations.


\begin{figure}
 \centering
 \includegraphics[width=\linewidth]{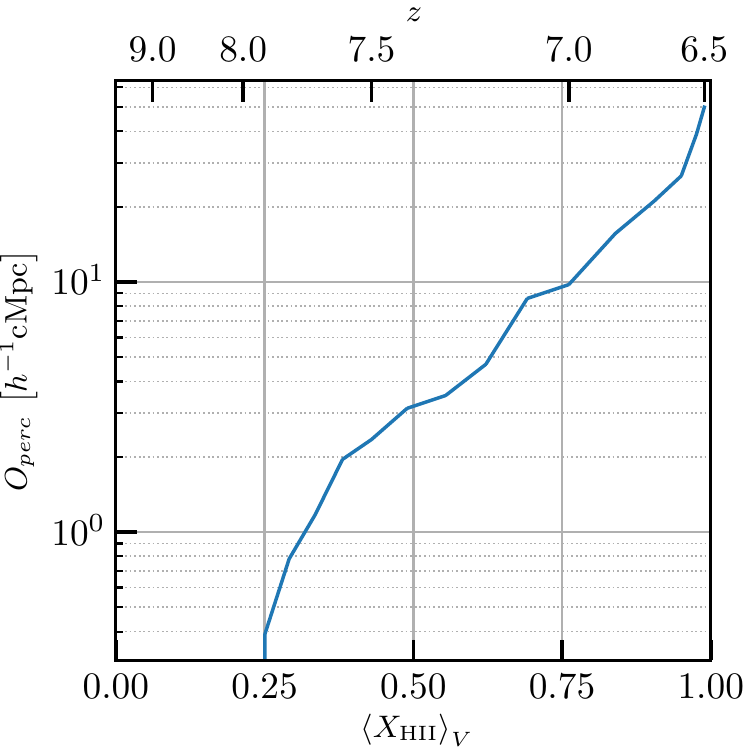}
 \caption{Maximum value of $O$ in fully ionized regions ($\xhii>0.99$) above which the regions still percolate as a function of volume weighted ionization fraction $\left<X_{\mathrm{HII}}\right>_V$ (lower axis) or redshift $z$ (upper axis). The curve refers to the stellar sources simulation.}\label{fig:reion_perc}
\end{figure}

We define the maximum radius of percolation, $O_{perc}$, as the maximum value of $O$, so that for regions larger than $O_{perc}$ the connections are severed and no percolation occurs. In \autoref{fig:reion_perc} we plot $O_{perc}$ as a function of the volume averaged ionization fraction (or, equivalently, redshift). 
 The results are obtained from the stellar sources only simulation, but hold for \hii in all scenarios. Although it is possible to have multiple POs in a given three dimensional volume (unlike in two dimensions), we only ever observe one at a time. This is expected as the ionized bubbles are anchored on the single emerging cosmic web.

We see that percolation sets in at $\vmxhii=0.25$, i.e. $z=7.9$. This is consistent with the maximum number of independent bubbles observed at $z=8$ (cf. Section~\ref{sec:bnum_res}), before further mergers decrease their numbers and increase their size, leading to percolation. 
The percolation radius grows exponentially with the volume averaged ionization fraction until it approaches the box size (corresponding to a radius of \SI{50}{\cMpch}, cf. Section~\ref{sec:gran_res}) as a natural barrier.


\begin{figure}
 \centering
 \includegraphics[width=\linewidth]{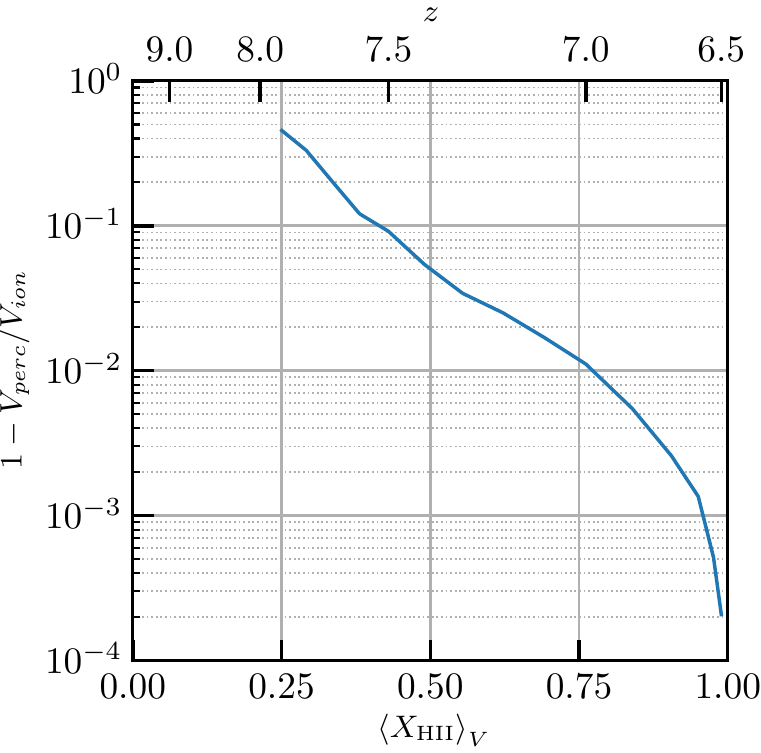}
 \caption{Fraction of the total fully ionized volume $V_{ion}$ not in the percolating object as a function of volume weighted average ionization fraction $\left<X_{\mathrm{HII}}\right>_V$ (lower axis) or redshift $z$ (upper axis). The curve refer to the stellar sources simulation.
}\label{fig:perc_vol_rat}
\end{figure}

\begin{figure}
 \centering
 \includegraphics[width=\linewidth]{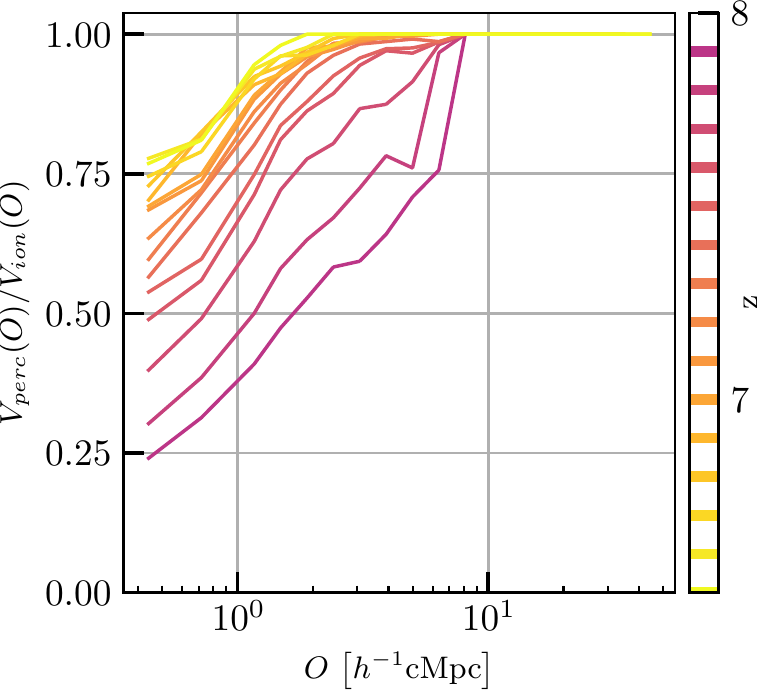}
 \caption{Fraction of the fully ionized volume at a given opening value $O$ in the percolating object at various redshifts in the range $6.5\leq z \leq 7.9$ (i.e. from its emergence to the end of reionization). The curve refer to the stellar sources simulation.}\label{fig:perc_rad_vol}
\end{figure}

In \autoref{fig:perc_vol_rat} we show how much of the fully ionized volume is (or is not) contained within the PO once it emerges.
When percolation first appears at $\vmxhii=0.25$ the PO contains $54.6\%$ of the total ionized volume, while the rest resides in independent bubbles. At the half-point of reionization the latter encompass only 5\% of the ionized volume, a percentage that drops to 0.02\% towards the end of reionization.

To understand how the sizes of the ionized regions inside and outside the PO differ,  in \autoref{fig:perc_rad_vol} we show the ratio between the volume within the PO and the total fully ionized volume at a given opening value. The most striking feature observed is the universal scale of \SI{\sim 8}{\cMpch} above which all volume resides in the PO. From this follows that, above such radius, the volume distribution between the ionized regions forming the PO is identical to the one shown in \autoref{fig:gran_z}. For smaller radii the volume distribution inside the PO is skewed towards larger values of $O$. The overlap behaviour agrees also very well with the typical bubble size from \ref{sec:gran_res}.

Initially, the PO contains 75\% of the volume with $O\approx\SI{5}{\cMpch}$ and only $\sim25\%$ in the smallest regions with $O\leq\SI{1}{\cMpch}$. Still, the PO only encompasses the previously mentioned 54.6\% of the total ionized volume, as these small regions dominate the total ionized volume at the time of its emergence (see Section~\ref{sec:gran_res}). Subsequently, the volume share in the PO approaches unity for all opening radii, while the fraction of the total ionized volume in small regions drops (see \autoref{fig:gran_z}). The combination of these two effects leads to the quick rise in the volume fraction of the PO as shown in \autoref{fig:perc_vol_rat}.

At the end of reionization all volume with $O\geq\SI{2}{\cMpch}$ is found in the PO and only 25\% of the volume in the smallest ionized regions remains disconnected. This is indicative of the ongoing formation of new ionized regions in underdense areas, far from the main drivers of reionization. So, although 75\% of the ionized volume with small opening value is connected to the PO, possibly residing in surface features, there still is an appreciable volume of independent bubbles of small radius.

\section{Connection to Underlying Fields}\label{sec:physfields}

While we formed an understanding of the population statistics of the ionized regions, we have so far neglected their connection to the underlying large scale structure in which they are formed. In this section we will therefore determine how ionized and neutral regions are related to the density field and to the ionizing sources that shape them. We note that the box size of our simulations  limits somewhat the significance of our results due to cosmic variance and exclusion of larger modes. For simplicity and conciseness we restrict ourselves to fully ionized HII regions ($\xhii>0.99$).

\subsection{Density Field}\label{sec:dens_res}

As the sources of radiation form in correspondence of density peaks, a strong connection between ionized bubbles and matter overdensity is expected. In this section we will quantify this assumption by evaluating the cross-correlation between the distribution of the structural centres and the density field. Unlike previous authors (e.g. \citealt{shin_cosmological_2008, xu_h_2019}) we are in the unique position to be able to use the structural centers of the ionized regions instead of just the ionized volume as a whole, allowing a much more accurate representation of the correlations (see \autoref{sec:mat_xcorr}) as the volume of an extended ionized region is dominated by its outskirts.

\subsubsection{Density-Centre Cross-Correlation}\label{sec:mat_xcorr}

\begin{figure*}
 \includegraphics[width=\textwidth]{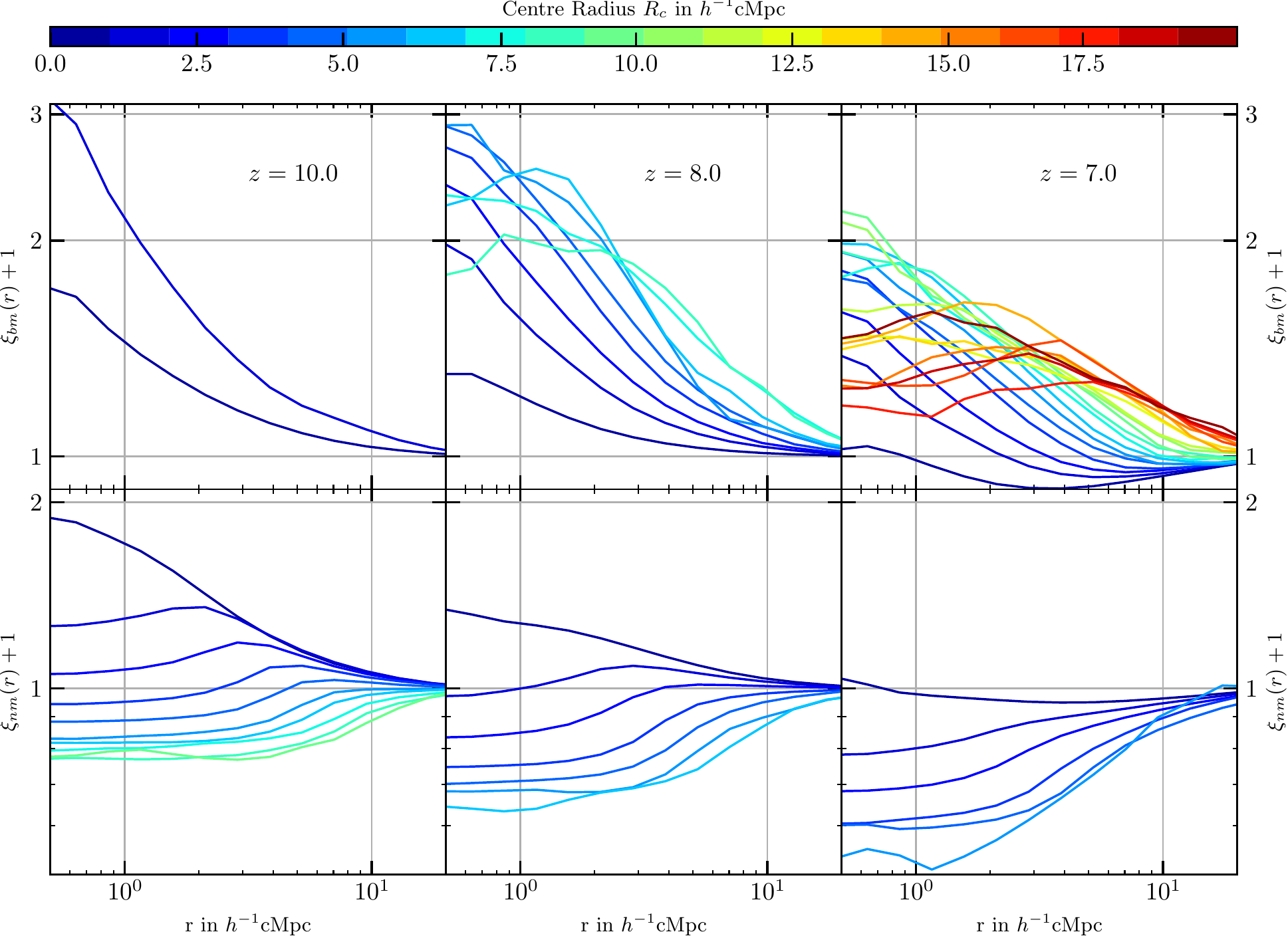}
 \caption{Cross-correlation between the matter density and the structural centres of ionized regions $\xi_{bm}$ (upper panels) and of neutral regions $\xi_{nm}$  (lower) at $z=10$ (left), 8 (middle) and 7 (right) as a function of separation $r$ between structuring element centres of radius $R_c$ and matter. For visibility we apply an offset of one.
}\label{fig:corrs}
\end{figure*}

To understand how ionized regions of different dimensions are connected to their large scale density environment, we discuss the evolution of the cross-correlation between their structural centres and matter density, defined as: 
\begin{align}
 \xi_{bm}(r) &= \mathcal{F}^{-1}\left(\left|\mathcal{F}(\delta_m)\mathcal{F}^\dagger(\frac{w_c}{\bar w_c}-1)\right|^2\right),
\end{align}
where $r$ is a given separation, $\delta_m$ is the matter overdensity, $w_c$ is the centre weight, $\mathcal{F}$ is the (discrete) Fourier transform and $\mathcal{F}^{-1}$ its inverse. 

The cross-correlation evaluated in 20 bins of centre radius, $R_c$, is shown in the upper panels of \autoref{fig:corrs} at $z={10,8,7}$. At all times we find that small ionized regions exhibit a centrally peaked correlation with the density. At the highest redshifts, the strength of the correlation increases with $R_c$, so that larger ionized regions are also in denser large scale environments. At $z=8$, though, while the above trend is maintained on large scales, this is not the case at small scales, where the correlations flatten for $R_c\gtrsim\SI{6}{\cMpch}$ and the strength of the correlation actually decreases with increasing $R_c$. As reionization proceeds, this behaviour is generally maintained  and the flattening is shifted towards higher values of $R_c$, for $R_c\gtrsim\SI{11}{\cMpch}$. On large scales the ordering with $R_c$ remains, and we also observe that these centrally peaked regions are becoming increasingly anti-correlated to the large scale matter distribution with decreasing $R_c$.

Similarly, we evaluate the cross-correlations with neutral rather than ionized regions, $\xi_{nm}$, and plot them in the lower panels of \autoref{fig:corrs}. Also in this case we have a clear trend with $R_c$, just in an inverted order, which stays the same at all times and radii, with the exception of the largest $R_c$ bin, the one most afflicted by cosmic variance. We find that at $z=10$ the smallest neutral regions are indeed exclusively overdense and only at $R_c\gtrsim\SI{3}{\cMpch}$ regions start to become centrally underdense, while only beyond $R_c\gtrsim\SI{5}{\cMpch}$ do they reside in large scale underdense regions. As reionization proceeds, they become less and less dense at all separations until at $z=7$ only the smallest $R_c$ bin is slightly overdense in the very centre.

The increased correlation with increasing $R_c$ for the centrally peaked profiles of the ionized regions is readily explained by the luminosity of the sources driving their growth. To form a larger ionized bubble at a given time the sources must have emitted more ionizing photons, which is the case for rare density peaks residing in overdense regions. The decrease of $\xi_{bm}$ with redshift for a given $R_c$ bin also follows form this picture, as at later times less rare peaks in lower density environments had the time to form ionized bubbles of the same size as those of the rarer peaks at earlier times. We interpret the flattened cross-correlation profiles as indicative of the formation of super bubbles. In fact, once the overlapping ionized regions close any remaining hole between them, a large ionized region emerges that is not anymore centred on a central source. The conditions for this to happen early in the reionization process are only met in environments where enough sources reside in close proximity, and each of which can contribute a significant ionized bubble. This can occur only in large scale environments even more exceptional than those hosting single rare peaks and therefore the large scale density is even more enhanced.

Also the cross-correlations of the neutral regions are related to the location of the ionized bubbles. A small neutral region needs to be closely surrounded by ionized regions, which implies that they can only be found in large scale overdense environs. Large neutral regions  with $R_c\gtrsim\SI{5}{\cMpch}$ inhibit the large scale underdense regions of our model universe. As reionization proceeds, the neutral regions are driven back into the underdense parts of the IGM, until only these can  host significant neutral gas towards the end of reionization.

\subsubsection{Bias of Bubble Centres}\label{sec:mat_bias}

An important parameter describing the cross-correlation between bubble centres and matter density field is the linear bias factor of the bubbles, $b = \xi_{bm}/\xi_{mm}$, where $\xi_{mm}$ is the matter auto-correlation. We calculate the bias parameter of neutral regions equivalently by just replacing $\xi_{bm}$ with $\xi_{nm}$. We evaluate $b$ following \cite{gao_assembly_2007} and the modifications in \cite{busch_tlt1_2019}, allowing for negative bias values. We calculate $b$ in four logarithmic radial bins from \SI{6}{\Mpch} to \SI{20}{\Mpch} and find the best common value via least squares. We restrict the analysis to samples with at least 100 centres.

\begin{figure}
 \centering
 \includegraphics[width=\linewidth]{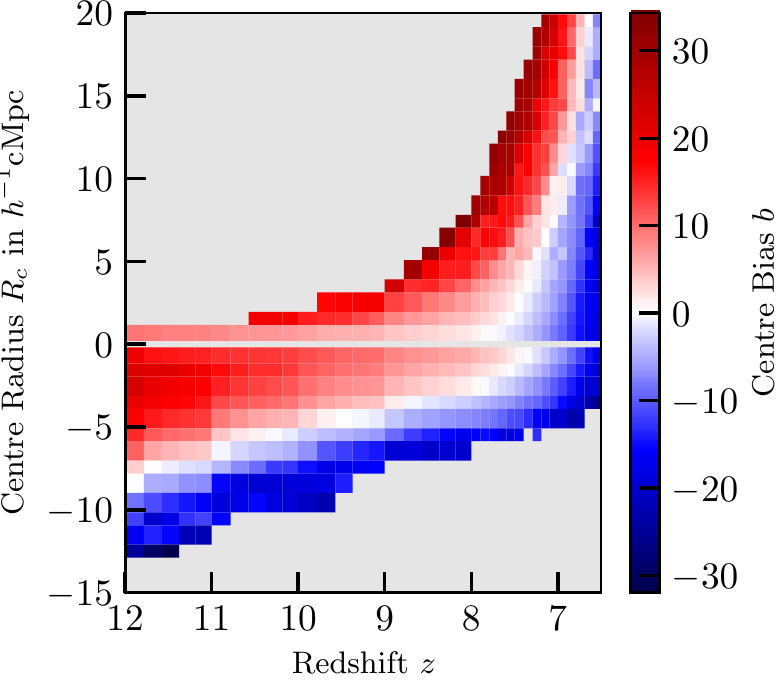}
 \caption{Bias of bubble centres, $b$, as a function of centre radius, $R_c$, and redshift, $z$, measured on scales from $6$ to \SI{20}{\cMpch}. We only show samples with at least 100 centres.
 Positive (negative) values of $R_c$ refer to ionized (neutral) regions.}\label{fig:bias}
\end{figure}

The combined results for ionized and neutral regions is shown in \autoref{fig:bias} where, following the convention established in \autoref{sec:of_intro}, the regions have, respectively, positive and negative values of $R_c$. The colour indicates the bias parameter for structural centres within a given $R_c$ bin at a given redshift. Red and blue indicate a positive and negative bias, corresponding to overdense and underdense environments, respectively. White marks the transition region of null bias, i.e. environments at the cosmic mean density. 

Consistently with the behaviour of the large scale cross-correlation discussed earlier, here we find that, at a fixed $R_c$, the bias decreases with decreasing redshift, for both ionized and neutral regions. For the smallest ionized bubbles such decrease leads to a null bias shortly after $z=8$, and by $z \sim 7$ regions with $R_c\lesssim\SI{10}{\cMpch}$ are already anti-correlated with matter.
The distribution in \autoref{fig:bias} shows that the bubbles are extremely biased tracers of the density field. Bubbles of all sizes appear first in unusually overdense regions (marked in red), and the bias of these newly formed bubbles increases with $R_c$ up to a maximum of $b\gtrsim 30$ at redshift $z\sim 8$, and then it remains roughly constant at $b \sim 30$ at lower redshifts.

Also the neutral regions show a strong bias evolution with redshift (i.e. a continuous decrease), being the mirror image of the ionized regions: at early times we find positive bias for small regions and negative bias for large ones, while at late times there are only small neutral regions with negative bias. The highest bias value is now associated to small ($\sim\SI{2}{\cMpch}$) neutral regions at the earliest redshift.

The behaviour observed in \autoref{fig:bias} can be explained by the arguments given in the previous section, i.e. a transition from single sources driven bubbles to very large ionized regions resulting from the merging of multiple bubbles. The decrease in bias values is a response to the merging of ionized regions in high density environments, that prevents the existence of bubbles with small $R_c$. These can only exist in the outskirts of the large ionized regions or in isolation in very underdense environments that are ionized last.

The imprint of an ``inside-out'' reionization scenario can be seen also on the bias of the centres of neutral regions. As already mentioned, initially small neutral regions exist only between ionized bubbles which lie sufficiently close to each other, i.e. only in peculiar environments, therefore inducing bias values larger than those of ionized regions at the same time. In contrast, the large neutral patches (i.e. which do not contain a single ionized cell) can exist only in very remote underdense environments. Towards the end of reionization all volume in overdense environments is ionized and only a few small neutral patches far from the sources prevail. With this the bias becomes uniformly negative.

There is little previous work on the bias of reionization bubbles. The different definitions of the correlated objects for which the bias is calculated makes it hard to compare results. \cite{shin_cosmological_2008} calculated the bias of any volume element above $\xhii=0.5$ and also found a bias value greater than 1. Due to the volume weighting of their measurement, their value mostly characterised the furthest outlying parts of the ionized regions (half of the volume in a sphere lies in the outermost 21\% of the radius). They also did not split their bubbles by size, but used the total population, corresponding to a volume weighted mean of our values (i.e. folding the volume distribution of \autoref{fig:gran_z} column-wise into \autoref{fig:bias}). It is therefore understandable that their values differ strongly from ours and are hardly comparable. \cite{xu_h_2019} extended this study by additionally computing the bias of the 21 cm differential brightness temperature using an advanced physical model but still used the whole field in both cases, still preventing a direct comparison in terms of numerical values of the bias. At least qualitatively we agree with their findings of an avoidance between neutral hydrogen and matter overdensity.



In order to get more representative results, we plan to extend these investigations to larger boxes that suffer less from cosmic variance and missing large scale modes.

\subsection{Luminosity Field}\label{sec:lum_res}

The driver behind hydrogen reionization is the luminosity originating from stellar type sources residing in galaxies. Here we investigate in more detail this connection by cross-correlating the ionization and luminosity fields. As expected, luminosity follows density in a biased fashion, confirming the results from the previous sections, obtained from the less noisy density field.

\begin{figure*}
 \includegraphics[width=\textwidth]{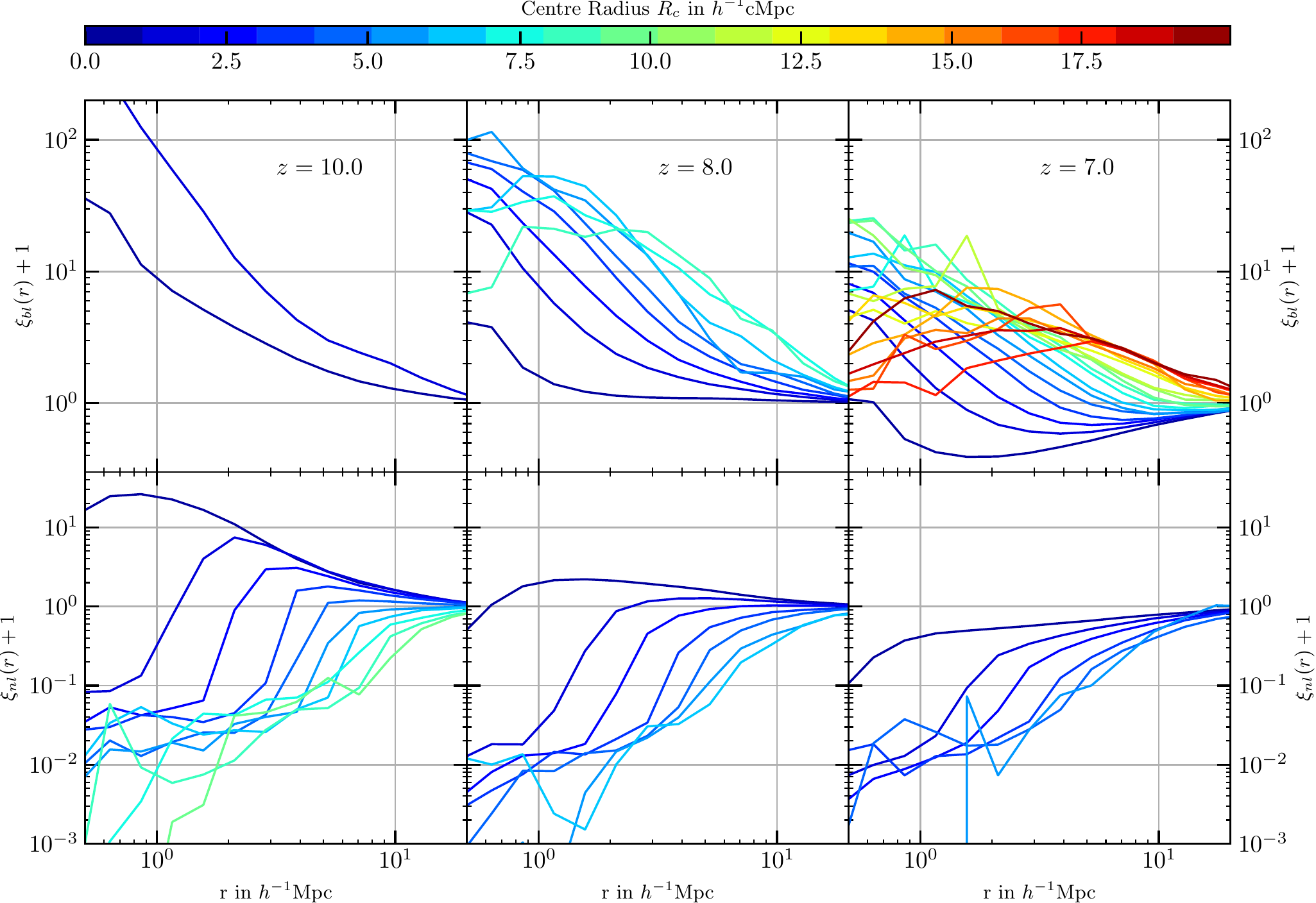}
 \caption{Cross-correlation between the luminosity density and the structural centres of ionized regions $\xi_{bl}$ (upper panels) and of neutral regions $\xi_{nl}$  (lower) at $z=10$ (left), 8 (middle) and 7 (right) as a function of separation $r$ between structuring element centres of radius $R_c$ and matter. For visibility we apply an offset of one.
 }\label{fig:lum_xcorr}
\end{figure*}

Similarly to what was done in \autoref{sec:mat_xcorr}, we cross-correlate the weighted centres of the bubbles with the luminosity field, normalised to the mean luminosity at any given redshift, and show it in \autoref{fig:lum_xcorr}.
As expected, the results are very similar to those discussed in \autoref{sec:dens_res}, especially on large scales where the luminosity field is just tracing the matter density field, albeit with a bias. Thus, the discussion on the behaviour at large scales in  \autoref{sec:dens_res} applies also here. At small scales, though, we observe an enhancement, indicative of a strong scale dependent luminosity bias for ionized, and even more for neutral (shown in the lower panels),  regions.
In fact, while for the centres of ionized regions the scale dependence seems to consist only of an increment in magnitude with decreasing separation $r$, for neutral regions the change is more dramatic, as even the sign changes in some cases. While small neutral regions at $z=10$ are positively correlated with matter on all scales, we find that for all but the smallest bin in $R_c$, $\xi_{nl}$ becomes negative on the scale of the neutral region, even for the two bins that had uniformly positive $\xi_{nm}$ values. This continues at later times, were we see a much stronger decrease in $\xi_{nl}$ than in $\xi_{nm}$.

The luminosity correlations in \autoref{fig:lum_xcorr} show clearly the selection on exceptionally luminous and faint areas for the structural centres of ionized and neutral regions, respectively. These areas are special in that their properties deviate strongly from the average bias of luminosity with respect to matter. While on large scales $\xi_{bl}$ and $\xi_{bm}$, and $\xi_{nl}$ and $\xi_{nm}$ just differ by a constant factor as expected from linear bias, the small scales behave clearly differently. Ionized regions are centred on especially luminous overdensities, giving $\xi_{bl}$ an extra boost over $\xi_{bm}$ on small scales. On the contrary, for small neutral regions positive correlations in $\xi_{nm}$ correspond to negative $\xi_{nl}$ values early during the initial phases of reionization, as these pockets of gas do not get ionized by nearby sources and lack local luminosity.

\section{Conclusions}\label{sec:conclusions}
 
We used a binary representation of the ionization fraction fields of hydrogen and helium to investigate the morphology of the ionized bubbles produced by radiative transfer simulations of cosmic reionization covering different source scenarios (\citealt{eide_epoch_2018}, \citealt{eide_next_2020}). For this we transformed the binary images with the Euclidean distance transform (EDT) and the morphological opening transform using a series of spherical structuring elements of increasing radius. The resulting opening field (OF) provides the spatial distribution of the local size of the bubbles. Additionally, the combination of these two transforms allows us to deconstruct the bubbles into a minimal set of overlapping, maximally large spherical structuring elements whose centre distribution we call the minimal centre set (MCS). Measuring the volume of the regions, density distributions and their connectivity at different radii offers insight into the bubble sizes and arrangements. The same process is applied to neutral regions as the complement of the ionized bubbles. Furthermore we can use the centers of the bubbles to find cross correlations with the density and luminosity fields to estimate typical profiles and bias values of the centres of ionized bubbles and neutral regions during the Epoch of Reionization (EoR).


A general finding of the present study is that, with the ionization threshold of 0.99 chosen to construct the binary images, there are no morphological indicators of the hydrogen ionization fraction that let us reliably distinguish between the different source scenarios, as expected for a stellar dominated reionization model, in which the physical conditions of the fully H-ionized regions are determined by stars \citep{eide_epoch_2018,eide_next_2020}. 
For example, we find no difference in either totally ionized volume or number of bubbles, apart from an increase in \heiii-regions at late times in scenarios including hard sources. However, these regions are still so small that we cannot say much about their morphology given our resolution.
These results are expected to change when looking at the 21cm images obtained from the different source scenarios, as the harder radiation emitted by ISM, XRBs and BHs changes the thermal state of the neutral and partially ionized IGM between bubbles.


In the range $6.5\leq z \leq 10$, the volume fraction of totally ionized cells increases exponentially, with decreasing redshift with a slight flattening when approaching unity between $z=7$ and $z=6$. At the same time we find that the connected ionized regions shift in their size distribution from a few cells and typical radii below \SI{1}{\cMpch} to a few \si{\cMpc} just before percolation sets in during the height
of reionization ($7\leq z \leq 8$), when most of the volume of the universe gets ionized. At the end of this process the volume is dominated by ionized regions with tens of \si{\cMpc} radius. It is also at this point that the first resolved \heiii-regions appear in the presence of thermal ISM emission and nuclear accreting BHs, although they do not outgrow \SI{1}{\cMpch}.

The volume in bubbles of different scales is distributed over a varying number of bubbles. Their size distribution changes from a shape reminiscent of the mass function of the haloes that form them   (a power-law with exponential cut-off at the upper end) at early times, to a single connected region on the largest scales and just a few disconnected smaller regions, well after percolation. During the time frame investigated here, \heiii remains in the first stage, as the bubbles are still very much connected with the local drivers.

To find the time of the percolation process following the bubble overlap and the scales of the resulting percolating object (PO), we search for connected regions above a given radius. We find that percolation first occurs at $z\gtrsim8$, i.e. $\vmxhii\approx0.25$. The PO rapidly expands its radius so that  just after $z\approx7.4$, at $\vmxhii=0.5$, we already find a PO made up of regions with at least \SI{3}{\cMpch}. At this point only $5\%$ of the fully ionized volume is not contained within the PO. Those separate ionized regions are of ever decreasing radius, with a maximum value of only $\sim \SI{2}{\cMpch}$ at $z=6.5$. Bubbles with radii above $\sim\SI{8}{\cMpch}$ are never found outside the PO.


We quantify the connection between bubbles and the cosmic density field by looking at the cross-correlations between the structural centres of ionized regions and neutral regions, and find that the results can be surprising. At early times ($z\approx 10$), the smallest neutral patches are actually in regions more overdense than their ionized counterparts. This can be explained by the fact that in order to restrict a neutral region to such a small size, it must be confined by a number of ionized regions, which happens only in a very overdense environment. Larger neutral regions, instead, tend to be centred on underdensities, where the appearance of ionized regions limiting their size is exceedingly unlikely. As the dimension of the neutral regions decreases with time, so does their average central density. Newly appearing larger ionized regions generally increase in average central overdensity with increasing size, while for a given size this quantity decreases.  This regularity is broken during the later stages of reionization, at $z \approx 7$, as we find a few more peculiarities. Not only can small ionized regions be found solely in large-scale underdense environments, but also there is a decrease in density towards the center of the largest ionized bubbles (which is already observed at $z \approx 8$ for the largest ionized regions). We interpret this as the result of the merger of formerly separated smaller bubbles centred on high overdensities that then appear in the outskirts of large ionized regions, while smaller bubbles can only evade merging if they are far enough removed from the bulk of the ionized volume.

We use the bubble-matter cross-correlation functions to calculate for the first time the linear bias values for the bubble centers. We find the largest bubbles at a given time to be extremely biased with respect to matter, with values of $10\lesssim b \lesssim 35$. This is a result of the exceptional circumstances that are required to form the largest ionized regions, as only multiple neighbouring bubbles of strong sources who merge are able to produce them. This is supported by the fact that the maximum bias value at a given time increases towards the point of percolation and subsequently decreases again, when the effect of radiation is increasingly unlocalised due to the dramatically increased mean free path. Just as we already saw in the correlation functions for the central values, we also find on large scales that small bubbles are avoiding matter, which leads to negative bias values. It is only in very underdense regions that they are not loosening their tight connection to matter density peaks due to merging.

The application of this novel approach to spatially resolved quantitative morphology is not limited to the use case of reionization. Other problems the method naturally lends itself to include the morphology of large scale structure as identified in galaxy surveys and Lyman-$\alpha$ tomography, but also smaller scale problems such as star forming filaments and other structures within galaxies themselves.

To conclude, our new morphological description opens up a new perspective on the local structure of reionization which can be directly connected to other physical quantities, as will be further explored in future publications.

\section*{Acknowledgements}
 

We would like to thank Fredrik Meyer for his helpful comments on the notation in this paper.

\section*{Data Availability}

The data underlying this article will be shared on reasonable request to the corresponding author.



\bibliography{phd} 
\bibliographystyle{mnras}

\begin{appendices}
 
\section{Fourier Space Dilation}\label{sec:four_dil}

The scale and resolution of our simulations dictate the use of rather large structuring elements. A simple direct-comparison approach for the morphological erosion and dilation operations scales as $\mathcal{O}(N\cdot M)$, where $N$ is the number of simulation cells and $M$ the number of cells in the structuring elements. $M$ scales as $O^3$ with the opening value for spherical structuring elements and therefore becomes large very quickly.
To speed up the operation, we reimplement the opening procedure and use a fast Fourier transform (FFT) based approach following \citet{kosheleva_fast_1997}, which therefore has a scaling of $\mathcal{O}(N\log N)$. This approach uses two facts:
\begin{enumerate}
 \item Morphological dilation $\oplus$ can be implemented as a convolution operation between the structuring element $S$ and the binary field (BF) $X$, which in turn can be implemented as a multiplication in Fourier space.
 \item The erosion of a BF is just the negation of the dilation of the negation of the same BF: $X \ominus S = \neg ( \neg X \oplus S )$.
\end{enumerate}
As we only implement a new dilation operation, the opening now becomes
\begin{equation}
 X\circ S = (\neg ( \neg X \oplus S ))\oplus S.
\end{equation}

While the new approach can slow the calculation for small structuring elements by orders of magnitude, it also decreases the run time by similar factors for large structuring elements. For simplicity we used the Fourier approach for all structuring element radii, as most of our cases contained many radii beyond $\sim10$ cells, which incidentally is roughly the point at which in our implementation the direct approach becomes slower.

We point the reader to \autoref{sec:edt_intro} for an approach to calculating the opening field in an even more time-saving manner if the Euclidean distance transform is also to be calculated.

\section{Formal description of the Minimal Centre Set and the Bubble Tree}
\label{app:bubble}

In \autoref{sec:min_struc} we only described the basic principle of the construction of the minimal centre set (MCS) and append a more detailed description here.

We first identify all cells with a specific opening value $O'$. We then construct a kD-tree from the positions of all cells for which $\lfloor E \rfloor = O'$ in units of the cell size $l_c$. All these positions can host a structuring element (SE) with radius $R=O'$ when using integer opening steps in units of $l_c$. For other step sizes we would have to adapt this criterion in order to just select those positions that fall into the range of EDT values corresponding to the desired value $O'$.

We use this kD-tree to find the candidate SE centre that lies closest to a given cell with $O=O'$. We then count the number of cells for which the given candidate centre lies closest, and call the count for each candidate centre $N_c$. If $N_c$ is at least one, we know that this is the centre of an essential structuring element. For most candidate centres $N_c$ is zero and they can therefore be neglected when one wants to reconstruct the binary image.

We then calculate a weighting factor $w_c$ that expresses how much of the structuring element is actually needed to reconstruct the opening field. For this we divide the cell count $N_c$ by the number of cells in the structuring element with $R=O'$, $N_{sph,O'}$:
\begin{equation}
 w_c = \frac{N_c}{N_{sph,O'}}.
\end{equation}
Cells that do not host a centre of an essential bubble have a weight of $0$ while all essential centres will have a weight in the range $0<w_c\leq 1$. 

Formally spoken, the tree of a bubble is rooted in the sphere centre at $\v x_R$ of each region of constant opening value that does not lie within a region of larger opening value. We can express this as:
\begin{align}
 w_c(\v x_R) &> 0\\
 \lceil E(\v x_R) \rceil &= O(\v x_R).
\end{align}
A spherical region is the child of one with a larger opening value if it is centred on a point $\v x_C$ within this larger region:
\begin{align}
 w_c(\v x_C) &> 0\\
 \lceil E(\v x_C) \rceil &< O(\v x_C).
\end{align}

The above construction leads to the classification of every single cell region as an independent bubble. To avoid this, we choose to attribute the cells among these which share a face with a cell of larger opening value as children to the centre of this larger region. Whether or not these bubbles are truly independent according to the above definition cannot be decided, as it can not be discerned whether they are overlapping bubbles (having a radius equal to the local opening value) or just surface features (when their centre lies within a region of larger opening value). As we expect the latter case to occur more often  for regions of opening value of one cell that are attached to larger regions, we artificially impose this condition. Failing to do so would lead to a strong artificial bias of the smallest regions lying on the surface of larger ones. 
Isolated one-cell regions are still fully considered.

With the above prescription every ionized cell in the binary image and every centre in the MCS belong exactly to one bubble tree. The construction is therefore unique for a given binary field.

For an illustration of the concepts laid out in detail above and the results for a realistic example in 2D we refer the reader to \autoref{fig:mcs_weighting} and \autoref{fig:mcs_example} in \autoref{sec:min_struc}.

Incidentally, this is closely related to performing a watershed transform on the EDT, an approach previously taken by \cite{lin_distribution_2016}. Our approach differs in details and in that we still retain the information of the distribution of scales within the EDT watershed regions. Furthermore we also record how the watershed regions are connected with each other. This places them in a greater context of complexly connected ionized regions, information that is lost if one only considers each bubble individually. For the most part our bubble definition should yield almost identical results as \cite{lin_distribution_2016}, modulo implementation details of the watershed transform and resolution of the grid. One important systematic difference is the treatment of a smaller SE center that resides in the cap of an intersection further from the centre of the larger SE in the intersection, as is the case for E in relation to B in \autoref{fig:mcs_weighting}. As the minimum of the EDT lies on the plane of intersection separating the two SE caps, the pure EDT watershed of \cite{lin_distribution_2016} would find two bubbles while we find one bubble as the centre of the smaller SE lies within the larger SE.

\end{appendices}

\label{lastpage}
\end{document}